\documentstyle[epsf,12pt]{article}
\def\bc{\begin{center}}
\def\ec{\end{center}}
\def\be{\begin{equation}}
\def\ee{\end{equation}}
\def\bea{\begin{eqnarray}}
\def\eea{\end{eqnarray}}
\def\nn{\nonumber}
\def\ltap{\ \raisebox{-.4ex}{\rlap{$\sim$}} \raisebox{.4ex}{$<$}\ }
\def\gtap{\ \raisebox{-.4ex}{\rlap{$\sim$}} \raisebox{.4ex}{$>$}\ }
\begin{document}
\pagestyle{empty} 
\vspace{-0.6in}
\begin{flushright}
CERN-TH/97-30 \\
ROME 97/1168 \\
ROM2F /97/09 
\end{flushright}
\vskip 0.2 cm
\centerline{\Large{\bf{CHARMING PENGUINS IN \boldmath$B$ DECAYS}}}
\vskip 1.4cm
\centerline{M.~Ciuchini$^{1,\star}$, E.~Franco$^2$,
G. Martinelli$^{2}$, L.~Silvestrini$^3$}
\centerline{$^1$  Theory Division, CERN, 1211 Geneva 23, Switzerland.}
\centerline{$^2$ Dip. di Fisica, Univ. ``La Sapienza"  and INFN,}
\centerline{Sezione di Roma, P.le A. Moro, I-00185 Rome, Italy.}
\centerline{$^3$ Dip. di Fisica, Univ. di Roma ``Tor Vergata''
and INFN,}
\centerline{Sezione di Roma II, 
Via della Ricerca Scientifica 1, I-00133 Rome, Italy.}
\abstract{Full expressions of the $B^0_d \to  \pi^+ \pi^-$
and $B^0_d \to  \pi^0 \pi^0$ amplitudes,  given in terms of matrix elements
of   operators of the effective weak Hamiltonian, are used to  study the
 dependence of the relevant branching ratios on the different
contributions.  The  uncertainty in the extraction of
the  weak phase $\alpha$ from the measurement of the
time-dependent asymmetry in   $B^0_d \to  \pi^+ \pi^-$ decays is also 
 analyzed.
We find that, among  several effects which may  enhance the
$B^0_d \to  \pi^0 \pi^0$ branching ratio,  the most important is  due
to  ``charming penguin" diagrams that  have never been studied before.
These diagrams  easily increase   $BR(B^0_d \to \pi^0 \pi^0)$ 
up to  a value of $1$--$3 \times 10^{-6}$.
The same effect produces, however, a large error in the extraction of $\alpha$
from the measurement of the $B^0_d \to  \pi^+ \pi^-$ time-dependent asymmetry.
We show that it is possible  to  determine charming-penguin amplitudes
from  the experimental measurement  of many decay rates. 
Their effect is impressive in
$B^+ \to \pi^+ K^0$ and  $B^0_d \to K^+ \pi^-$ decays, where
charming-penguin contributions easily give values of
$BR(B^+ \to \pi^+ K^0)$ and $BR(B^0_d \to K^+ \pi^-)$ of  about 
$1 \times 10^{-5}$.  Among  other possibilities, 
we also suggest to use    $B^0_d  \to  K^0 \bar K^0$,
the $BR$ of which can be as large as $2$--$3 \times10^{-6}$,
to determine the size of charming-penguin amplitudes.} 
\vskip 5 cm
\centerline{$^\star$ On leave of absence from INFN, Sezione Sanit\`a,
 V.le Regina Elena 299, Rome, Italy.}
\vfill\eject
\pagestyle{empty}\clearpage
\setcounter{page}{1}
\pagestyle{plain}
\newpage 
\pagestyle{plain} \setcounter{page}{1}
 
\section{Introduction} \label{intro}
The study of  $B^0_d \to \pi \pi$ ($\bar B^0_d \to \pi \pi$)
decays is of paramount importance
for our understanding of CP-violation in the Standard Model and
beyond.   In particular the  measurement of the time-dependent
asymmetry 
\bea 
{\cal A}(t) &=&\frac{N(B^0_d \to \pi^+ \pi^-)(t)- N(\bar B^0_d \to \pi^+ \pi^-)
(t)}{N(B^0_d \to \pi^+ \pi^-)(t)+ N(\bar B^0_d \to \pi^+ \pi^-)
(t)}\nn \\
&=& \frac{(1-\vert \lambda\vert^2) \cos (\Delta M_d t )
-2 Im \lambda \sin (\Delta M_d t )}{1+\vert \lambda\vert^2} \label{eq:asy}
\eea
may allow the extraction of the CP-violating phase  $\alpha$, see for
example~\cite{nirs}.  The cleanest  method to
extract $\sin 2 \alpha$ is from the measurement of  the asymmetry, combined
with the separate determination of the   $I=0$ and $I=2$ decay amplitudes,
including  the relative phase~\cite{gronau}. These   
can be obtained by measuring  the $B^+ \to  \pi^+ \pi^0$,
$B^0_d \to \pi^+ \pi^-$ and $B^0_d \to \pi^0 \pi^0$ (and the 
corresponding ones in the   $\bar B^0_d$ case) branching ratios.
With these measurements, we  get rid of our ignorance of the
hadronic matrix elements of the weak Hamiltonian.
 Unfortunately, most of the theoretical analyses tend to predict
a very small  $B^0_d \to \pi^0 \pi^0$ branching ratio, thus making the
model-independent extraction of $\sin 2 \alpha$ impossible in practice.
\par If $\sin 2 \alpha$ has to be extracted  from $B^0_d \to \pi^+ \pi^-$ only,
the main uncertainty comes from  the contribution proportional
to $\lambda_t= V_{td} V^\star_{tb}$, which is usually called ``penguin
pollution".  In several studies, the decay rates and the uncertainty
of $\sin 2 \alpha$  have been estimated by using  some specific model to
evaluate the hadronic matrix elements of the four-fermion operators
entering  the effective weak Hamiltonian~\cite{dibart}--\cite{alek}.
 In the most popular approaches
the amplitudes have been computed by assuming the factorization
hypothesis.
 The matrix elements of the weak currents necessary for the evaluation
 of the factorized amplitudes  are then  taken from a 
specific quark model or from the HQET~\cite{formq}--\cite{forml}.  
\par In this paper, we present a ``model-independent" analysis 
of the uncertainty on $\sin 2 \alpha$ and of the ratio
$R=\Gamma(B_d^0 \to \pi^0\pi^0)/\Gamma(B_d^0 \to \pi^+\pi^-)$.
 By ``model-independent" we mean that we
do not make specific assumptions on the hadronic matrix elements of
 the operators,
such as factorization or the absence of final state interactions (FSI).
On the basis of simple ``qualitative" physical considerations, we allow, 
instead,
the matrix elements to vary within certain ``reasonable" ranges, and
check the stability of the results against such variations.  
This is particularly relevant for $R$, because of the delicate
cancellations occurring 
between different amplitudes present in the $B_d^0 \to \pi^0\pi^0$ case.
Indeed, for this decay, the assumption of factorization and 
of the absence of FSI,
or any approximation used to predict the value of the amplitude,
may lead  to an underestimate of  the value of the decay rate. 
\par Our calculations  are  based on
complete expressions of the decays amplitudes  for 
$B^+ \to  \pi^+ \pi^0$,
$B^0_d \to \pi^+ \pi^-$ and $B^0_d \to \pi^0 \pi^0$,
given in terms of diagrams
representing   Wick contractions of the operators of
 the effective Hamiltonian between the relevant external states.
These formulae allow  us to  clarify  assumptions and approximations usually
made to evaluate the amplitudes, which have not been spelt explicitly
in previous studies.  In particular, we show the presence
of  diagrams,  involving operators containing  charmed quarks 
(defined as $Q_1$ and $Q_2$ in sec.~\ref{sec:formulae}) that  contribute
to  the penguin pollution, and that have never been considered before. We call
these diagrams ``charming penguins".
\par Among  several effects which are able to enhance the
$B^0_d \to  \pi^0 \pi^0$ branching ratio,  the most remarkable is due
precisely to  charming  penguins.
Their contribution  may increase the estimate
of  $BR(B^0_d \to \pi^0 \pi^0)$ up to  a value of $1$--$3 \times 10^{-6}$. 
The reason is that, unlike the 
case of the penguin operators $Q_3$--$Q_{10}$ which have small
Wilson coefficients (of
 order $\alpha_s /12 \pi \ln (m_t^2/\mu^2)$), the  coefficients of
$Q_1$ and $Q_2$ are of $O(1)$ and  there is no reason to believe 
the corresponding matrix elements to be small.
Charming  penguins  are also relevant
for the  $B^0_d \to \pi^+ \pi^-$ amplitude and may give a large shift 
$\Delta=\sin 2 \alpha^{eff}-\sin 2 \alpha \sim 0.4$--$0.8$ 
 between  the  physical value of  $\sin 2 \alpha$
 and  the ``effective" value, $\sin 2 \alpha^{eff}$, 
which can be extracted from the experimental measurement
of $Im \lambda$.
As a comparison, when charming  penguins are not included, the
typical value is $\Delta\sim 0.1$. 
As $\Delta$ increases, however, also
$BR(B^0_d \to \pi^0 \pi^0)$ becomes larger, thus opening the possibility
of extracting $\sin 2\alpha$ with the isospin analysis
proposed in ref.~\cite{gronau}.
\par We finally show that many decay rates
are expected to be dominated by  charming-penguin diagrams.   Among
the various possibilities, we consider  $B^0_d  \to  K^0 \bar K^0$,
$B^+ \to \pi^+ K^0$ and $B^0_d \to K^+ \pi^-$   decays.
In these cases, we give  explicit formulae for the  amplitudes,
show that the largest contributions are those expected  from charming penguins
and estimate the corresponding branching ratios.
\par The most impressive effect of charming  penguins is found in
$B^+ \to \pi^+ K^0$ and $B^0_d \to K^+ \pi^-$   decays.
Assuming reasonable values for  the charming-penguin contributions,
we find that their
branching ratios may even become  larger than  $BR(B^0_d \to \pi^+ \pi^-)$. 
This observation  is particularly
interesting because, in absence  of  charming-penguin diagrams,
the  $B^+ \to \pi^+ K^0$ and $B^0_d \to K^+ \pi^-$ rates 
turn out to be  very small either because there is 
a  Cabibbo suppression or because the non-Cabibbo suppressed terms
 come from penguin operators which have rather small Wilson coefficients
 (unless the corresponding matrix elements are exceedingly large).
While finishing this analysis, we were informed that the CLEO collaboration
has measured $BR(B^0_d \to K^+ \pi^-)= (1.5 ^{+0.5}_{-0.4} \pm 0.2 ) 
\times 10^{-5}$~\cite{CLEO}.
The prediction that charming-penguin diagrams are important and give large 
$B^+ \to \pi^+ K^0$ and $B^0_d \to K^+ \pi^-$  decay rates
is supported by this measurement.
By using the experimental information, we predict
$BR(B^+ \to \pi^+ K^0)\sim 1\times 10^{-5}$ and we call for a search of
this decay mode.
\par
Other interesting decay channels, where charming penguins are expected 
to  play an important role, such  as $B^0_d \to \pi^0 \eta$
(or $\eta^\prime$), $B^0_d \to \phi K^0$ or  $B^0_d \to \eta \eta$ decays,
will be 
extensively discussed elsewhere~\cite{cfms2}. 
\par The plan of this paper is the following. In section~\ref{sec:formulae},
we introduce  the effective Hamiltonian
given in terms of four-fermion operators, and of the corresponding 
Wilson coefficients; we also define the full set of diagrams in terms
of which  the $B \to  \pi \pi$ amplitudes can be expressed; 
the final formulae of the different amplitudes are given  at the end of
 this section. 
Formulae and approximations for the $B^0_d  \to  K^0 \bar K^0$,
$B^+ \to \pi^+ K^0$ and $B^0_d \to K^+ \pi^-$  amplitudes are discussed
 in section~\ref{sec:altri}. In section~\ref{sec:me}, we present several 
physical  arguments which are used to guide us in estimating
the matrix elements; we also explain the main
criteria used in the numerical analysis. In section~\ref{sec:results}, 
we give  and discuss the main
numerical results  for $R$, $\sin 2 \alpha$ and for
the  $B^0_d  \to  K^0 \bar K^0$,
$B^+ \to \pi^+ K^0$ and $B^0_d \to K^+ \pi^-$ branching ratios.
\section{Relevant formulae for \boldmath$B \to \pi \pi$ decays}
\label{sec:formulae}
The effective weak Hamiltonian relevant for $B \to \pi \pi$ decays
is given by 
\bea
{\cal H}_{eff}^{\Delta B=1}&=&\lambda_u \frac {G_F} {\sqrt{2}}
\Bigl[  \Bigl( C_1(\mu)\left( Q^u_1(\mu) - Q_1(\mu) \right) +
C_2(\mu)\left( Q^u_2(\mu) - Q_2(\mu) \right)  \Bigr)\nn\\
&+&\tau \, \vec C(\mu)\cdot\vec Q(\mu)   \Bigr]
\protect\label{eh}
\eea
where $\vec Q(\mu)=(Q_1(\mu),Q_2(\mu),\dots, Q_{10}(\mu))$,
$\vec C(\mu)=(C_1(\mu),C_2(\mu),\dots, C_{10}(\mu))$,
$\lambda_u = V_{ud} V^\star_{ub}$ and similarly
$\lambda_c$ and $\lambda_t$; $\tau=-\lambda_t/\lambda_u$
 and $\mu$ is the renormalization scale of the operators $Q_i$.
A convenient  basis of operators~\cite{russi}--\cite{lus},
 when QCD and QED corrections are taken
into account, is 
\bea
Q^u_{ 1}&=&({\bar b}d)_{ (V-A)}
    ({\bar u}u)_{ (V-A)}
   \nn\\
Q^u_{ 2}&=&({\bar b}u)_{ (V-A)}
    ({\bar u}d)_{ (V-A)}
\nn \\
Q_{ 1}&=&({\bar b}d)_{ (V-A)}
    ({\bar c}c)_{ (V-A)}
\nn \\
Q_{ 2}&=&({\bar b}c)_{ (V-A)}
    ({\bar c}d)_{ (V-A)} \nn \\
Q_{ 3,5} &=& ({\bar b}d)_{ (V-A)}
    \sum_{q}({\bar q}q)_{ (V\mp A)}
\nn \\
Q_{ 4} &=&\sum_{q} ({\bar b}q)_{ (V-A)}
    ({\bar q}d)_{ (V - A)}
\protect\label{eq:basis}\\
Q_{6} &=& -2 \sum_{q} ({\bar b}q)_{ (S+P)}
    ({\bar q}d)_{ (S-P)}
\nn \\
Q_{ 7,9} &=& \frac{3}{2}({\bar b}d)_
    { (V-A)}\sum_{q}e_{ q}({\bar q}q)_
    { (V\pm A)} \nn \\
Q_{8} &=& - 3\sum_{q}  e_q({\bar b}q)_
    { (S+P)} ({\bar q}d)_{ (S-P)} \nn \\
Q_{ 10} &=& \frac{3}{2}\sum_{q}e_{ q}({\bar b}q)_
{ (V-A)}({\bar q}d)_{ (V -  A)} \nn     
\eea
where the subscripts $(V \pm A)$ and $(S \pm P)$ 
indicate the chiral structures and $e_q$ denotes the quark electric charge
($e_u=2/3$, $e_d=-1/3$, etc.).
The sum over the quarks $q$  runs over the active flavours at the
scale $\mu$.
\begin{figure}   
    \centering
    \epsfxsize=0.50\textwidth
    \leavevmode\epsffile{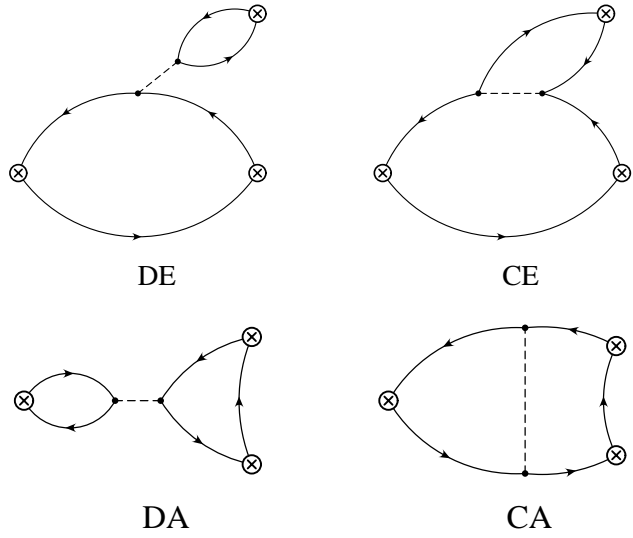}
       \caption[]{\it{Non-penguin diagrams.
The dashed line represents the four-fermion operator.}}
    \protect\label{fig:first}
\end{figure} 
\begin{figure}   
    \centering
    \epsfxsize=0.50\textwidth
    \leavevmode\epsffile{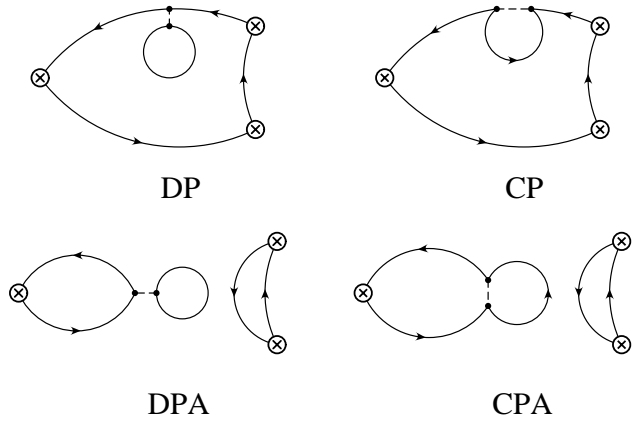}
       \caption[]{\it{Penguin diagrams.}}
    \protect\label{fig:last}
\end{figure} 
\par Wick contractions of ${\cal H}_{eff}$ between hadronic states
give rise to the diagrams  shown in 
figs.~\ref{fig:first}--\ref{fig:last}: these are ``Disconnected Emission"
($DE$), denoted  also as $T$ or $T^\prime$ in ref.~\cite{gronau1}; 
the colour suppressed (non-factorizable)
``Connected Emission" ($CE$),  denoted also as $C$ or $C^\prime$;
``Disconnected Annihilation" ($DA$), denoted also as $A$ or $A^\prime$;
``Connected Annihilation" ($CA$), denoted also as $E$ or $E^\prime$;
``Disconnected Penguin" ($DP$),
``Connected  Penguin" ($CP$), ``Disconnected Penguin Annihilation" ($DPA$),
``Connected  Penguin Annihilation" ($CPA$).  
We assume $SU(2)$ isospin symmetry. For penguin
diagrams, we introduce a label which identifies the quark flavour  in the 
penguin loop. Thus, for example, $DP(s)$ denotes the disconnected penguin
diagram $DP$ of fig.~\ref{fig:last}, with a strange quark in the
internal  loop
(a similar notation is adopted also for  $CP$, $DPA$ and $CPA$).
 Since we do not distinguish up and down quarks, we simply call
$DP$ ($CP$, $DPA$ and $CPA$)  those diagrams with an up or a down quark  in the loop.
In eq.~(\ref{eq:basis}), different Dirac structures appear, namely 
$L =(V - A) \times (V-A)$,  $ R=(V - A) \times (V+ A)$ and
$S= (S+P) \times (S-P)$~\footnote{ By Fierz rearrangement we could
have adopted a basis where only $(V - A) \times (V\pm A)$ structures
are present. We prefer however to work with operators written
in terms of colour-singlet bilinears.}. 
Thus, for example,  the notation $CE_L$ and
$CE_S$  denote the connected emission diagrams with a $(V - A) \times (V-A)$
or $(S+P) \times (S-P)$ operator inserted, respectively.
We are now ready to give the complete expressions for the decay
amplitudes under study.
It is convenient to write 
their $\Delta I=3/2$ and $\Delta I=1/2$ components separately
\bea A^{+0}&=&
A(B^+ \to \pi^+ \pi^0)=\frac{G_F}{\sqrt{2}} \lambda_u \, \frac{3}{\sqrt{2}} A_2 
\ ,\label{eq:ap0}\\
A^{+-}&=&A(B^0_d \to \pi^+ \pi^-)= \frac{G_F}{\sqrt{2}} \lambda_u \, (A_2-A_0) \ , \label{eq:apm} \\
A^{00}&=&A(B^0_d \to \pi^0 \pi^0)= \frac{G_F}{\sqrt{2}} \lambda_u \, 
(\sqrt{2} A_2 + \frac{1}{\sqrt{2}} A_0) 
\ , \label{eq:a00} \eea 
where 
\be
A_2= A^u_2+ \tau  A^t_2  \ , \,\,\,\,\,\,\,\,\,\,\,\, A_0 = A^u_0+ \tau  A^t_0
\, ,\ee
and
\bea
A^u_2 &=& -\frac{1}{3} \left[C_1 +C_2\right] \left(DE_L+CE_L\right) 
\label{eq:mu}\\
A^t_2 &=& -\frac{1}{2}\left[
C_7 \left(DE_R+CE_R\right) -2 C_8 \left(DE_S+CE_S\right) \right. \nn
\\ &\,&+ \left. \left(C_9+C_{10}\right) \left(DE_L +CE_L\right) \right] \, ,
\label{eq:mostro2}\eea
\bea
A^u_0 &=& C_1  \left[-\frac{1}{3} DE_L+\frac{2}{3}CE_L +DA_L +
 \left(DP_L-DP_L(c)\right) + \left(DPA_L-DPA_L(c)\right)\right] \nn \\
&+& C_2 \left[\frac{2}{3} DE_L-\frac{1}{3}CE_L + CA_L +
 \left( CP_L-CP_L(c)\right) + \left( CPA_L-CPA_L(c)\right)\right]\label{eq:pippo}
 \eea \bea
A^t_0 &=&C_1  \left[ DP_L(c)+DPA_L(c) \right]
+C_2 \left[ CP_L(c)+CPA_L(c) \right]  \nn \\ 
&+& C_3  \left[ CE_L + CA_L + 2 DA_L + CP_L + 2 DP_L + DP_L(c) + DP_L(s)\right.
\nn\\
&+& \left. CPA_L + 2 DPA_L + DPA_L(c) + DPA_L(s) \right] \nn \\
&+& C_4 \left[ DE_L + 2 CA_L + DA_L + 2 CP_L + CP_L(c) + CP_L(s)
 + DP_L \right.\nn\\
&+& \left.  2 CPA_L + CPA_L(c) + CPA_L(s) + DPA_L \right] \nn \\
&+& C_5  \left[ CE_R + CA_R + 2 DA_R + CP_R + 2 DP_R + DP_R(c) + DP_R(s)  
\right. \nn \\
&+& \left.         CPA_R + 2 DPA_R + DPA_R(c) + DPA_R(s) \right] \nn \\
&-& 2 C_6 \left[ DE_S + 2 CA_S + DA_S + 2 CP_S + CP_S(c) + CP_S(s)
 + DP_S  \right. \label{eq:mostro}
 \\
&+&  \left.  2 CPA_S + CPA_S(c) + CPA_S(s) + DPA_S \right]  \nn \\
& +&  \frac{1}{2} C_7 \left[  CE_R - DE_R - CA_R + DA_R-
 CP_R + DP_R + 2 DP_R(c) - DP_R(s)\right.\nn\\
&-&\left.  CPA_R + DPA_R + 2 DPA_R(c) - DPA_R(s) 
\right] \nn \\
&+&  C_8 \left[ CE_S - DE_S - CA_S + DA_S- CP_S - 2 CP_S(c) + CP_S(s) + DP_S
  \right.\nn\\
&-& \left.  CPA_S - 2 CPA_S(c) + CPA_S(s) + DPA_S \right] \nn \\
&+&\frac{1}{2} C_9 \left[   CE_L - DE_L - CA_L + DA_L - CP_L + 
         DP_L + 2 DP_L(c) - DP_L(s) \right.\nn\\
&-& \left. CPA_L + DPA_L + 2 DPA_L(c) - DPA_L(s) 
\right] \nn \\   
&-& \frac{1}{2} C_{10} \left[ CE_L - DE_L - CA_L + DA_L - CP_L - 2 CP_L(c) 
+ CP_L(s)  + DP_L \right.\nn\\
&-& \left.   CPA_L - 2 CPA_L(c) + CPA_L(s) + DPA_L \right]  \nn
 \ .
\eea
In eqs.~(\ref{eq:mu})--(\ref{eq:mostro}) we have not shown explicitly the
 argument $\mu$
of the Wilson coefficients $C_1, \dots, C_{10}$.   Notice that
also the diagrams are $\mu$-dependent, since they correspond to contractions
of  renormalized operators $Q_i(\mu)$.
For the fields we have assumed the standard convention $B^+=\bar b u$,
$B^0_d=\bar b d$, $\pi^+=u \bar d$, $\pi^-=-d \bar u$ and $\pi^0=-
1/\sqrt{2}(u \bar u- d \bar d)$.
\par In $A_0^u$,  penguin diagrams always  appear in the GIM combinations
$DP_L-DP_L(c)$, $DPA_L-DPA_L(c)$, $\dots$ (in the following
we denote these combinations as GIM-penguins).  Had we taken   $m_t \ll M_W$ and a  renormalization scale $\mu$ 
larger than the top quark mass, i.e. $m_t \ll  \mu \ll M_W$,
 similar GIM combinations would have  appeared in  $A^t_0$,
namely  $DP_L(t)-DP_L(c)$, etc. Since  the physical
value of $m_t$ is so large,  the diagrams  $DP_L(t)$, $CP_L(t)$,
  etc. are replaced by  complicated structures. These  arise from the
contractions of the penguin and electro-penguin  operators
$Q_3$--$Q_{10}$, originated in ${\cal H}_{eff}$  when we remove
the top quark from the effective theory. In the literature,   $A^t_0$ only
is   identified as penguin pollution.  We want to stress again  that in the
effective theory there are  ``penguin operators" $Q_3$--$Q_{10}$, 
which originate from the imperfect GIM cancellation occurring when
$\mu \ll m_t$ and ``penguin diagrams" which arise from the Wick
contractions of {\sl all} the operators of ${\cal H}_{eff}$.
 \par The coefficients of
the penguin operators  $Q_3$--$Q_{10}$ are of
 order $\alpha_s /12 \pi \ln (m_t^2/\mu^2)$ and contain
 the short distance contribution  (from  scales between $\mu$ and
 $m_t$) of the virtual top  and charm quarks. For  $\mu \sim m_b$, these coefficients
  are rather small,
 e.g. the dominant term is due to $Q_6$, for which $C_6/C_2 \sim -0.03$.
 Thus, unless the corresponding matrix elements are very large, 
 the penguin operators  are not
 expected to give large corrections, at least in the $B^0_d \to \pi^+ \pi^-$
 case.  In the $B^0_d \to \pi^0 \pi^0$ case, the relative correction due
 to penguin operators, $Q_6$ in particular, may be more important
 due to the large cancellations present in the amplitude (the so called
 ``colour suppression" to be discussed below).
 \par This is not the end of the story, however. In $A_0^t$  there are 
 other terms, specifically those in the first line
 of eq.~(\ref{eq:mostro}).  These terms, denoted as ``charming  penguins",
  come from   penguin contractions of the 
 operators $Q_1$ and $Q_2$, and have to be understood as long distance
 contributions in the matrix elements of these operators. 
 Because of the unitarity relation
 $\lambda_c=-\lambda_u-\lambda_t$, they give contributions  to both $A_0^u$, in the GIM combination
 $DP_L$-$DP_L(c)$, and to $A_0^t$.     
 Unlike the case
of the penguin operators $Q_3$--$Q_{10}$,  the  coefficients of
$Q_1$ and $Q_2$ are of $O(1)$ and, since there is no reason to believe 
the corresponding matrix elements to be 
small~\footnote{ This is true 
in spite of the fact that, for
 $(V - A) \times (V-A)$ operators, penguin diagrams 
vanish if we assume factorization, as demonstrated
by the important role that they are expected to play in
kaon decays.},  they are potentially relevant
even for the  $B^0_d \to \pi^+ \pi^-$ amplitude. 
 On the other hand We notice that their 
contribution can be further enhanced by the factor $\vert \tau \vert\sim 2$.
\par Penguin diagrams 
are also present in $A^u_0$ and can alter in $A^{+-}$ and $A^{00}$
the relative size of the term proportional to $\lambda_u$ 
with respect to that proportional to $\lambda_t$.  
This can be seen by looking to the expression of 
$\lambda$ introduced in eq.~(\ref{eq:asy}) in terms of the different amplitudes
\be
\lambda = \frac{q}{p} \frac{\overline{A}^{+-}}{A^{+-}}
= \frac{q}{p}    \frac{\lambda_u^\star}{\lambda_u}
\frac{1+\tau^\star (A^t_2-A^t_0)/(A^u_2-A^u_0)}
{1+\tau (A^t_2-A^t_0)/(A^u_2-A^u_0)} \ ,
\ee
where 
\bea \frac{q}{p} &=& \frac{V_{td}V^\star_{tb}}{V^\star_{td}V_{tb}}=
\frac{1-\rho-i \eta}{1-\rho+i \eta} = e^{-2i\beta} \nn \\
\frac{\lambda_u^\star}{\lambda_u}&=& \frac{V_{ub}V^\star_{ud}}{V_{ud}
V^\star_{ub}}=
\frac{\rho-i \eta}{\rho+i \eta} = e^{-2i\gamma} \nn \\
\tau&=& -\frac{1-\rho-i \eta}{\rho+i \eta} \label{eq:wolf} \eea
$\rho$ and $\eta$ being the CKM parameters in the Wolfenstein parametrization
\cite{wol} and $\beta +\gamma=\pi -\alpha$.
\par At this point we have all the formulae needed for the study of
the uncertainties on  $\sin 2 \alpha$ and in the calculation of $R$.
\section{\boldmath$B^0_d \to K^0 \bar K^0$, \boldmath$B^+ \to  \pi^+ K^0$
and $B^0_d \to K^+ \pi^-$}
\label{sec:altri}
\par We now show that the $B^0_d \to K^0 \bar K^0$,
 $B^+ \to  \pi^+ K^0$ and  $B^0_d \to K^+ \pi^-$   amplitudes  are  dominated
by GIM- and charming-penguin diagrams.  Let us consider
 the three cases separately. For simplicity,
 all the formulae in this section are computed 
 in the $SU(3)$ symmetric limit,
e.g. $DP_L(s)=DP_L$.
\vskip 0.2 cm \underline{$B^0_d \to K^0 \bar K^0$}
We start by studying this case for which all
the relevant quantities have already been defined. The amplitude
is dominated by GIM and charming-penguin diagrams
because
emission and annihilation  diagrams are only produced  by the insertion  of
the operators  $Q_3$--$Q_{10}$ which have very  small Wilson coefficients.
By defining  $A(B^0_d \to K^0 \bar K^0)=\lambda_u  {G_F}/ {\sqrt{2}}
\times  A_K$,  the complete expression of $A_K$ is given by
\bea
A_K &=&C_1  \left[ \left(DP_L- DP_L(c)\right) +\tau DP_L(c) + 
\left(DPA - 
DPA_L(c)\right) +\tau DPA_L(c) \right] \nn \\
&+&C_2 \left[  \left(CP_L- CP_L(c)\right) + \tau CP_L(c) + 
\left(CPA - CPA_L(c)\right)
+ \tau  CPA_L(c) \right]  \nn \\ 
&+& \tau \left\{ C_3  \left[ CE_L + CA_L + 2 DA_L + CP_L + 3 DP_L + DP_L(c) 
   \right.\right.
\nn\\
&+& \left. CPA_L + 3 DPA_L + DPA_L(c)  \right] \nn \\
&+& C_4 \left[ DE_L + 2 CA_L + DA_L + 3 CP_L + CP_L(c) 
 + DP_L \right.\nn\\
&+& \left.  3 CPA_L + CPA_L(c)  + DPA_L \right] \nn \\
&+& C_5  \left[ CE_R + CA_R + 2 DA_R + CP_R + 3 DP_R + DP_R(c)  
\right. \nn \\
&+& \left.         CPA_R + 3 DPA_R + DPA_R(c)  \right] \nn \\
&-& 2 C_6 \left[ DE_S + 2 CA_S + DA_S + 3 CP_S + CP_S(c) 
 + DP_S  \right.  \nn
 \\
&+&  \left.  3 CPA_S + CPA_S(c)  + DPA_S \right]  
\label{eq:mostriciattolo} \\
& +&  \frac{1}{2} C_7 \left[  -CE_R  - CA_R - 2 DA_R-
 CP_R  + 2 DP_R(c) \right.\nn\\
&-&\left.  CPA_R + 2 DPA_R(c)  
\right] \nn \\
&+&  C_8 \left[ DE_S + 2 CA_S +  DA_S  - 2 CP_S(c) 
+ DP_S  \right.\nn\\
&-& \left.   2 CPA_S(c)  + DPA_S \right] \nn \\
&+&\frac{1}{2} C_9 \left[  - CE_L  - CA_L - 2  DA_L - CP_L  
          + 2 DP_L(c)  \right.\nn\\
&-& \left. CPA_L  + 2 DPA_L(c) 
\right] \nn \\   
&-& 
 \frac{1}{2} C_{10} \left[ DE_L +2  CA_L + DA_L  - 2 CP_L(c) 
  + DP_L \right.\nn\\
&-& \left.\left.    2 CPA_L(c)  + DPA_L \right] \right\} \nn
\, . \eea 
By neglecting all  penguin-operator contributions we get
\bea  A_K &=&C_1  \left[ \left(DP_L- DP_L(c)\right) +\tau DP_L(c) + 
\left(DPA - 
DPA_L(c)\right) +\tau DPA_L(c) \right] \nn \\
&+&C_2 \left[  \left(CP_L- CP_L(c)\right) + \tau CP_L(c) + 
\left(CPA - CPA_L(c)\right)
+ \tau  CPA_L(c) \right]  \, . \label{eq:apkkb}
\eea 
This shows that unless the matrix elements of penguin operators, $Q_6$
for example, are 
exceedingly large (much larger than their estimates in the factorization
hypothesis~\cite{alek}), 
the amplitude is dominated by GIM- and charming-penguin diagrams.
Equation~(\ref{eq:apkkb}) is only given  to display the relevant terms;
in all our numerical calculations we always used the complete expressions
for all the amplitudes.
\vskip 0.2 cm \underline{$B^+ \to  \pi^+ K^0$} 
In order to study $B^+ \to \pi^+ K^0$ and $B^0_d \to K^+ \pi^-$ 
we have to introduce new quantities. For these  decays, the operator
$Q^u_1$ is replaced by $Q^{\prime \, u}_1=
({\bar b}s)_{ (V-A)} ({\bar u}u)_{ (V-A)}$ with a Wilson 
coefficient $C_1^\prime=C_1$, and similarly for all the 
other operators in (\ref{eq:basis}): this comes from the fact
that  we are considering 
now $\Delta B=-\Delta S=1$ transitions instead of the  $\Delta B=\Delta D=1$ 
ones of  sec.~\ref{sec:formulae}. 
We also define  $\lambda^\prime_u =V_{us} V^\star_{ub}$ and similarly
$\lambda^\prime_c$ and $\lambda^\prime_t$:
$\lambda^\prime_u$ is of  $O(\lambda^4)$ whereas 
 $\lambda^\prime_c$ and  $\lambda^\prime_u$ are of 
$O(\lambda^2)$, where $\lambda\sim 0.22$  is the sine of the Cabibbo angle
in the Wolfenstein approximation. With the exception
of $\lambda^\prime_{u,c,t}$, we will omit the superscript
$^\prime$ for the rest of this section.  
We write 
\be A(B^+ \to  \pi^+ K^0)=\frac{G_F}{\sqrt{2}}
\Bigl( \lambda^\prime_u  A^u_+ + \lambda^\prime_c  A^c_+ +
\lambda^\prime_t  A^t_+ \Bigr)\, , \label{eq:ap} \ee
where  
\bea A^u_+ &=&C_1  \left[ CA_L +DP_L  \right]
+ C_2 \left[  DA_L + CP_L  \right]  \nn \\ 
A^c_+&=&  C_1   DP_L(c) +C_2 CP_L(c)  \nn \eea
\bea A^t_+&=& - C_3  \left[ CE_L + CA_L  + CP_L + 3 DP_L + DP_L(c) \right]
\nn \\ &-& C_4\left[ DE_L +  DA_L + 3 CP_L + CP_L(c)  + DP_L \right] 
\nn\\
&-& C_5\left[ CE_R + +CA_R+ CP_R + 3 DP_R  + DP_R(c) \right]
\nn \\
&+& 2 C_6 \left[ DE_S + DA_S + 3 CP_S + CP_S(c) +  + DP_S\right]  
\label{eq:mostriciattolo1} \\
&-&  \frac{1}{2} C_7 \left[  -CE_R  + 2 CA_R -
 CP_R  + 2 DP_R(c) \right] \nn \\
&-&  C_8 \left[ DE_S - 2  DA_S  - 2 CP_S(c) + DP_S \right] 
\nn \\
&-&\frac{1}{2} C_9 \left[  - CE_L  +2 CA_L - CP_L  +2 DP_L(c) \right] \nn \\   
&-& 
 \frac{1}{2} C_{10} \left[ -DE_L  + 2 DA_L + 2 CP_L(c) -DP_L \right]  \nn
\, . \eea 
By neglecting all  penguin-operator contributions  and Cabibbo
suppressed terms we get
\be A(B^+ \to  \pi^+ K^0)=\frac{G_F}{\sqrt{2}}
 \lambda^\prime_c  A^c_+ =\frac{G_F}{\sqrt{2}}
 \lambda^\prime_c \left[C_1  DP_L(c) + C_2 CP_L(c) \right] 
  \, , \label{eq:apkkb1} \ee
\vskip 0.2 cm \underline{$B^0_d \to  K^+ \pi^-$} 
We write 
\be A(B^0_d \to  K^+ \pi^)=\frac{G_F}{\sqrt{2}}
\Bigl( \lambda^\prime_u  A^u_0 + \lambda^\prime_c  A^c_0 +
\lambda^\prime_t  A^t_0 \Bigr)\, , \label{eq:ap2} \ee
where  
\bea A^u_0 &=&C_1  \left[ -CE_L  - DP_L  \right]
+C_2 \left[  - DE_L - CP_L  \right]  \nn \\ 
A^c_0&=& - C_1   DP_L(c) - C_2 CP_L(c)  \nn \eea \bea
A^t_0&=& - C_3  \left[ -CE_L - CA_L  - CP_L - 3 DP_L - DP_L(c) \right]
\nn \\ &-& C_4\left[ -DE_L -  DA_L - 3 CP_L - CP_L(c)  - DP_L \right] 
\nn\\
&-& C_5\left[- CE_R-CA_R - CP_R - 3 DP_R  - DP_R(c) \right]
\nn \\
&+& 2 C_6 \left[ -DE_S - DA_S - 3 CP_S - CP_S(c) 
 - DP_S\right]  
\label{eq:mostriciattolo2} \\
&-&  \frac{1}{2} C_7 \left[  -2 CE_R  + CA_R +
 CP_R  - 2 DP_R(c) \right] \nn \\
&-&  C_8 \left[ 2 DE_S -  DA_S   +2  CP_S(c) - DP_S\right] 
\nn \\
&-&\frac{1}{2} C_9 \left[  - 2 CE_L  + CA_L + CP_L  -2 DP_L(c)
 \right] \nn \\   
&-& 
 \frac{1}{2} C_{10} \left[ -2 DE_L  +  DA_L  - 2 CP_L(c) 
 + DP_L \right]  \nn
\, . \eea 
By neglecting all  penguin-operator contributions  and Cabibbo
suppressed terms we get
\be A(B^0_d \to  K^+ \pi^-)=\frac{G_F}{\sqrt{2}}
 \lambda^\prime_c  A^c_0 = - \frac{G_F}{\sqrt{2}}
 \lambda^\prime_c \left[C_1  DP_L(c) + C_2 CP_L(c) \right]
  \, , \label{eq:apkkb2}\ee
In $A(B^+ \to \pi^+ K^0)$ and $A(B^0_d \to K^+ \pi^-)$,
if  charming-penguin diagrams are very small  then  both Cabibbo suppressed
 contributions and terms due to penguin operators must
 be included in the calculation since they are of the same size. 
\section{Estimates of the diagrams}
\label{sec:me}
In this section, we discuss the criteria adopted to evaluate the diagrams
appearing in eqs.~(\ref{eq:mu})-(\ref{eq:mostro}) and in sec.~\ref{sec:altri}.
Notice that the value  of the diagrams is given only once that
the renormalization prescription (RP) and renormalization
scale $\mu$ of the operators have been
fixed. Under a change of RP or $\mu$  the values of the
diagrams must be  changed in such a way as to compensate the corresponding changes
of the Wilson coefficients $C_i$, thus giving the same physical predictions
up to and including next-to-leading logarithmic 
corrections~\cite{noi}--\cite{epp}~\footnote{ Notice that under a change of
the RP  or of $\mu$, contributions attributed to matrix elements of some
operators can go into the Wilson coefficients or the matrix elements
of others, and viceversa.}.
The state of the art in the calculation of the matrix
elements of the operators is such that, given the complexity of the expressions
in  eqs.~(\ref{eq:mu})--(\ref{eq:mostro}), this turns out to be 
impossible. For example,  factorized amplitudes are RP and scale independent,
being expressed in terms of physical quantities. Thus they cannot
compensate a variation of the coefficient functions. On the other
hand,  the possibility that lattice calculations   be able
to compute (\ref{eq:mu})--(\ref{eq:mostro}) with sufficient
accuracy  appears to be  rather remote. This is particularly true for
$A^{00}$ where delicate cancellations are likely to occur between  different
contributions, see also the discussion and the end of this
section and section~\ref{sec:results}.
In the following, in order to check the stability of the results,  at  
fixed values of the diagrams   we vary the Wilson coefficients
by changing RP and by taking $2$ GeV $\le \mu \le 10$ GeV.
\par We now discuss the assumptions made in the evaluations
of  the different diagrams:
\begin{enumerate}\item[1)] \underline{$DPA$ and $CPA$} These are Zweig  suppressed
diagrams which  we assume to give a  negligible contribution.
\item[2)] \underline{Electro-penguins} In order to monitor
the effects of the electro-penguins, we only consider the contributions
coming from the operators $Q_9$ and $Q_{10}$ since these operators have 
 coefficients much larger than $Q_7$ and $Q_8$. 
\item[3)] \underline{$DE_L$ and $CE_L$} In most of the theoretical
analyses, these diagrams give the largest contribution to 
the $B^0_d \to \pi \pi$ amplitudes.
If only emissions are present,  there are three independent quantities namely  
$\vert DE_L\vert$,  $\vert CE_L \vert$ and $arg(DE_L \times CE_L^\star)$.  Without loss of generality we can then
write $CE_L=\xi DE_L e^{i \delta_\xi}$. We vary   $0.0 \le \xi \le 0.5$:
 this range covers the
value preferred  by the analysis of  $D$-meson  two-body  non-leptonic decays, 
which suggests $\xi \sim 0$, and   the value derived from  $a_2/a_1$ extracted
from $B \to D \pi$ and $B \to D \rho$ decays~\cite{dibart,xiref}. 
Moreover it includes
the canonical value $\xi=1/N_c$ where $N_c$ is the number of colours.
On the basis of some estimates of FSI~\cite{don,pham}, we
 do not expect the relative phase $\delta_\xi$ to be larger than $\sim 0.5$,
at the energy scales  flowing into the pion system in $B$-decays.
We enlarged the range of $\delta_\xi$ up to $\sim 1$ in order to check which
 kind of effect could be produced by a large phase.
\item[4)] \underline{$DA_L$ and $CA_L$}  These are diagrams usually neglected 
since arguments can be made that, in the factorization hypothesis, 
they are suppressed by a factor $f_B/M_B$
(besides  colour suppression in  $CA_L$)~\cite{gronau1}. 
$f_B/M_B$ is related to the  $B$-meson wave function  in the origin.
A further suppression factor comes from the matrix element of the
divergence of the vector current which creates the pion pair.
In ref.~\cite{buccella}, however, a fit to two-body $D$-meson decays,
resulted in a non negligible value for the  annihilation diagrams,
corresponding to $DA_L/DE_L \sim 0.3 (m_s-m_d) /f_\pi$. It is not clear how
these results should be scaled to the $B$-meson case and for degenerate quark
masses. 
\par Rescattering
effects, which have been shown to persist even for large quark masses
\cite{don}, can also enhance the value of annihilation diagrams with respect
 to  factorization  estimates~\cite{zenc},
see also~\cite{bgr}.
As discussed in ref.~\cite{zenc}, emission
and annihilation diagrams are connected via FSI. For example, the $CA$-diagram
can be seen as a $DE$ followed by rescattering of the final states.
In the $1/N_c$ expansion, since the scattering
amplitude is of order $1/N_c$, this gives immediately the correct 
leading  dependence on the number of colours for $CA_L$. In fact
$DE_L \sim N_c^{1/2}$,  $CA_L \sim N_c^{-1/2}$  and
$CA_L \sim \eta_A DE_L$, where $\eta_A$ is proportional to the scattering 
amplitude, which is of $O(1/N_c)$. The only  potentially large
contribution from annihilation diagrams comes indeed from the term 
$\propto C_2\, CA_L$ in eq.~(\ref{eq:daimp}), since all the other annihilation
contributions (either $DA_L$ or $CA_L$)  have (much) smaller  coefficients. 
Moreover, several arguments can be made to show that the value
of $DA_L$ is expected to be at most of the size of $CA_L$ and its largest
contribution multiplies $C_1$ which is about $1/5$ of $C_2$.  
In our numerical analysis,
we have explicitly checked, by varying $DA_L$ between zero and
$CA_L$, that its effect is rather marginal. For this reason,
in sec.~\ref{sec:results} we only discuss the case with $DA_L=0$.
To take into account  rescattering  effects, we parametrize $CA_L$
as $CA_L=i DE_L \eta_A$.
 $\eta_A$ is a complex ``inelasticity"
coefficient, the absolute value of  which  has been estimated to be of the
order of some tenth~\cite{don,pham}.  In our numerical study we take
it  real  with
$0 \le \eta_A \le 0.5$~\footnote{ All estimates give 
 $Re \, \eta_A \gg Im \, \eta_A$~\cite{don,pham}.}. 
The same rescattering mechanism relates
disconnected and connected   diagrams in the penguin case, which 
is discussed below.
\item[5)] \underline{GIM-penguins} Penguin-like contractions appear both in 
$A_0^u$ and $A_0^t$. Here following we discuss the two cases separately.
In $A_0^u$ we always find the  combinations $DP_L-DP_L(c)$ and
$CP_L-CP_L(c)$ and one could argue that, because
of the large final state energy at disposal,  GIM cancellation makes these contributions
negligible. The GIM mechanism, however, is expected to be effective
only at short distances, i.e. when  a  high momentum flows in the penguin loop.
For low momenta, i.e. if we look to long-distance effects,
these diagrams can also be  interpreted as
emission diagrams followed by rescattering. For example,
\bea CP_L-CP_L(c) &\sim&  DE_L(B^0_d \to \pi \pi) S_0(\pi \pi\to \pi \pi)\nn \\
&\, &- DE_L(B^0_d \to D \, D) S_0(D \, D \to \pi \pi) \nn \\
 &\sim&  f_\pi q^\mu 
\langle \pi(\vec p_B-\vec q)\vert J^{ub}_\mu \vert B^0_d \rangle
S_0(\pi \pi\to \pi \pi)  \nn \\ &\,&
-f_D q^\mu \langle D (\vec p_B-\vec q)\vert J^{cb}_\mu \vert B^0_d \rangle
S_0(D\, D\to \pi \pi) \, , \label{eq:cpexa} \eea
where $S_0$ is the strong interaction  $S$-matrix and $J^{ub,cb}_\mu$ 
are the weak vector  currents.
Since $f_D/f_\pi \sim 1.5$~\cite{flynn} and the form factors relative to 
$\langle \pi(\vec p_B-\vec q)\vert J^{bu}_\mu \vert B^0_d \rangle$ are expected
to be  smaller than those relative to 
$\langle D (\vec p_B-\vec q)\vert J^{bc}_\mu \vert B^0_d 
\rangle$~\cite{formq}--\cite{forml},
by about a factor of $2$, $F^{cb}(M_D^2) \sim 2\,  F^{ub}(M_\pi^2)$, 
it is not clear how  effective is the GIM cancellation between the two
contributions. A large cancellation between the charm and up contributions
may still take place
if the relative factor 
$S_0(D\, D\to \pi \pi)/S_0(\pi \pi\to \pi \pi)$  compensates
the differences due to  phase space,  decay constants and
form factors~\footnote{ In this respect a combined measurement
of the  $ B \to \pi \pi$ and $B \to D \, D$ branching ratios
would be very interesting.}.
\par The discussion of rescattering effects for
 GIM-penguins strictly follows that  of the annihilation diagrams made in 4).
 For this reason,  given our ignorance of $S_0$,
we use the parametrization  $CP_L-CP_L(c)=i DE_L \eta_P$,
with  $0 \le \eta_P\le 0.5$,  and ignore the contribution from $DP_L-DP_L(c)$.
\item[6)] \underline{Other penguin diagrams} 
In $A_0^t$,   penguin contributions are not GIM suppressed.  Thus 
we expect that rescattering effects play a minor role.
Penguin contributions are of two kinds: either they correspond to the insertion
of  left-left operators, $Q_1$, $Q_2$, $Q_3$, $Q_4$, $Q_9$ and $Q_{10}$,
or they are given by the insertion of $Q_5$ and $Q_6$, for example
in $DP_R$ or $DP_S$. \par In the previous section, we have noticed that 
penguin contractions of the operators $Q_1$ and $Q_2$  can give 
large effects since the corresponding coefficients are of $O(1)$.
In our numerical analysis we  find  that  a modest relative phase
between $DP_L$ and $CP_L$ can  have dramatic effects. For this reason
 we  introduce two  parameters $\eta_{L}$ and $\delta_L$, by writing
$DP_L=\vert DE_L\vert \eta_{L}$ and $CP_L=
e^{i \delta_L} \vert CE_L\vert \eta_{L}$, with 
$0.0 \le \eta_L \le 1.2$ and $0 \le \delta_L \le 0.5$. The range of values of $\eta_L$
is dictated from the fact that there is no reason to expect very  large/small
values for these matrix elements, so we take them of the same order than the
corresponding emission diagrams. As for the phase, we 
limit the maximum of $\delta_L$ to $0.5$, as we did for $\delta_\xi$.
For simplicity, we take $\eta_L$  flavour independent, i.e.
we use the same value  of $\eta_L$ for $DP_L$, $DP_L(s)$ and $DP_L(c)$.
We do not connect $CP_L$ to $DE_L$, as we did for GIM-penguins,
 because, as said above,  we do not have to advocate 
long-distance effects coming from emissions followed by rescattering.
 In general, we should consider a complex value of $\eta_L$. We checked,
however, that the largest effects come from the relative phase
between $DP_L$ and $CP_L$ and for this reason, only the case with $\eta_L$ real
will be discussed in the following. 
\par 
Left-right penguin contributions only appear together with all the
other contractions of $Q_5$ and $Q_6$. The latter have the
same topology of the diagrams considered so far, but different
chiral structures.  Rather than introducing another set of free parameters
for the $(V - A) \times (V+ A)$ and $(S+P) \times (S-P)$ diagrams, 
we consider globally the matrix elements of the operators
$Q_5$ and $Q_6$ and write 
\be \langle \pi \pi \vert Q_5 \vert B^0_d \rangle = \eta_5 CE_L \, ,
\,\,\,\,\,\,\,\,\, 
 \langle \pi \pi \vert Q_6 \vert B^0_d \rangle = \eta_6 DE_L \, .
\label{eq:q56} \ee
In kaon decays, there is a common prejudice that the operators
$Q_5$ and $Q_6$  trigger the octet enhancement~\footnote{ In the kaon
case, if $\mu$ is larger than the charm mass,
which plays the role of the top quark mass for GIM effects,  the
operators $Q_5$ and $Q_6$ are hidden in the matrix elements of
$Q^u_1-Q_1$ and $Q^u_2-Q_2$ (with $b \rightarrow s$ in the
operators of the effective Hamiltonian) .}. If really the explanation of the enhancement relies
on the matrix elements of these operators (as also suggested by lattice 
calculations~\cite{gavela12}), then $\eta_{5,6}$ can be as large as $5$.
Since the kinematical configuration is so different in $B$-decays, due
to the large mass of the $b$-quark, and many of the arguments for the
enhancement are based on the chiral expansion, we do not really expect
 $\eta_{5,6}$ to be   as large as in the kaon case.
Since we cannot exclude, however,  values of $O(1)$,  we vary
$0.0 \le \eta_{5,6} \le 2.0$. 
\end{enumerate}
We summarize the discussion above, by giving the expressions
(\ref{eq:mu})--(\ref{eq:mostro}) in units of $DE_L$, and in terms of the
parameters $\xi$, $\delta_\xi$, $\delta_L$ and of the $\eta_i$s
\bea
A^u_2 &=& -\frac{1}{3} \left(C_1 +C_2\right) \left(1+\xi e^{i\delta_\xi}\right)
\label{eq:mu2} \\
A^t_2 &=& -\frac{1}{2}\left(C_9+C_{10}\right) \left(1+\xi e^{i\delta_\xi}\right) 
 \, ,
\label{eq:mostro2s}\eea
\bea
A^u_0 &=& C_1  \left[-\frac{1}{3} +\frac{2}{3} \xi e^{i\delta_\xi}\right] \nn \\
&+& C_2 \left[\frac{2}{3} -\frac{1}{3}  \xi e^{i\delta_\xi}  + i( \eta_A +
 \eta_P)  \right] \label{eq:daimp} \\
A^t_0 &=&  \left[ C_1  +C_2 \xi e^{i\delta_L} \right] \eta_L  
+ \nn \\ 
&+& C_3  \left[ \xi e^{i\delta_\xi} + i\eta_A
 + \left(4+  \xi e^{i\delta_L}\right) \eta_L  \right] \nn \\
&+& C_4 \left[ 1 + 2 i\eta_A + 
(1+ 4 \xi e^{i\delta_L} ) \eta_L  \right] \nn \\
&+& C_5 \xi e^{i\delta_\xi} \eta_5 +C_6 \eta_6 \nn \\
&+&\frac{1}{2} C_9 \left[  -1+  \xi e^{i\delta_\xi}  - i  
  \eta_A  + \left(2 -\xi e^{i\delta_L} \right)\eta_L  
  \right] \nn \\   
&-& \frac{1}{2} C_{10} \left[ -1+  \xi e^{i\delta_\xi}  - i  
  \eta_A  +\left(1-  2  \xi e^{i\delta_L} \right)\eta_L 
 \right]    \, .
\label{eq:mostros}
\eea
\par Before discussing the numerical results of our analysis, we want to
add some observations about colour-suppressed processes.
The starting point are eqs.~(\ref{eq:mu})--(\ref{eq:mostro})
[(\ref{eq:mu2})--(\ref{eq:mostros})], which contain
the expressions  of  the relevant amplitudes.
For the sake of illustration, let us consider first the case where only
$DE_L$ and $CE_L$ are non vanishing and we neglect penguin 
operator contributions. In this case, when we insert 
(\ref{eq:mu})--(\ref{eq:mostro}) in eqs.~(\ref{eq:apm}) and (\ref{eq:a00})
the following combinations occur
\bea A^{+-} &\propto & C_1 CE_L + C_2 DE_L  =
C_2 DE_L \left( 1 + \frac{C_1}{C_2} \xi e^{i\delta_\xi} \right)
\label{eq:xipm}\\
A^{00} &\propto & C_1 DE_L + C_2 CE_L =
 C_1 DE_L \left( 1 + \frac{C_2}{C_1} \xi e^{i\delta_\xi} \right)
 \label{eq:xi00} \, .\eea
Numerically, the ratio  $-C_1/C_2 \sim 0.2$--$0.3$ is approximatively
equal in size and opposite in sign to the expected  value of $\xi$.
This implies   that the second term in eq.~(\ref{eq:xipm}) is a small correction
(of the order of $10$\%) to the first one  while the two terms
in eq.~(\ref{eq:xi00}) tend to cancel
(for small values of the phase $\delta_\xi$). This, together
with the smallness of $C_1$,  is at the origin of
colour suppression.  In $D$-decays, where a very small value of $\xi$  is
preferred, colour suppression is expected to be less effective~\footnote{ This
 could alternatively mean that factorization is a poor approximation.}. 
 \par As can be read from eqs.~(\ref{eq:mu})--(\ref{eq:mostro}),
a similar colour-suppression pattern is  present in all the $\Delta I=1/2$
annihilation and penguin diagrams generated by the operators $Q_1$--$Q_2$,
which are the operators with the largest coefficients: 
\bea 
A^{00} &\propto & C_1  \left[  DA_L + \left(DP_L - DP_L(c) \right)
+ \tau DP_L(c) \right] \nn \\ &+& 
C_2  \left[  CA_L + \left(CP_L - CP_L(c) \right)
+ \tau CP_L(c) \right] + \dots  \label{eq:cs} \eea
Thus, if all
the connected diagrams are about a factor of $\xi$ smaller than the 
corresponding disconnected ones, it is very difficult
to resurrect the $B^0_d \to \pi^0 \pi^0$ amplitude, unless the 
matrix elements of the penguin operators $Q_3$--$Q_{10}$ are very large. 
The same arguments, however, could be applied to kaon decays, with the
surprising conclusion that $\vert A_0\vert  \sim 2
\vert  A_2\vert$ in sharp disagreement
with the experimental observation  $\vert A_0\vert \gg \vert A_2\vert$. 
It is then clear that dynamical effects play a fundamental role in this game. 
\par Of course, it is possible to argue that the dynamics of the
decay for kaons and $B$-mesons is completely different and that
factorization is a very bad approximation  for light mesons. It is
clear, however, that one cannot exclude some residual dynamical effect
which enhances the $B^0_d \to \pi^0 \pi^0$ amplitude over its factorized 
value. The contribution of $Q_1$ and $Q_2$ to $A^t_0$,
non-zero phases $\delta_\xi$ and $\delta_L$ or
the connected-disconnected diagram-exchange mechanism for $CA_L$ and $CP_L
-CP_L(c)$ discussed above  may provide such  effects. 
\section{Numerical results}
\label{sec:results}
In this section, we present  numerical results for
the quantities $\sin 2 \alpha$ and  $R$, that only depend 
on ratios of amplitudes.  The latter are 
all proportional, via the parameters
$\xi$, $\dots$, $\eta_6$, to  $DE_L$, the value of which cancels out 
in the ratios. \par
One could, however,  also be  interested to know the variation of 
$BR(B^0_d \to \pi^+ \pi^-)$ for different assumptions about GIM-penguin,
charming-penguin, annihilation diagrams etc.
Thus we also give below $R_{+ -}=BR(B^0_d \to \pi^+ \pi^-)/
BR(B^0_d \to \pi^+ \pi^-)\vert_{DE_L}$, 
where $BR(B^0_d \to \pi^+ \pi^-)$
is the branching ratio computed for a given set of the parameters
$\xi$, $\dots$, $\eta_6$, whilst  $BR(B^0_d \to \pi^+ \pi^-)\vert_{DE_L}$
is the branching ratio with all the diagrams put to zero, but $DE_L$.
In this way, the reader can use his preferred model to compute
$DE_L$ and predict the physical value of  $BR(B^0_d \to \pi^+ \pi^-)$.
In the following, $R$ will also denote $BR(B^0_d \to \pi^0 \pi^0)/
BR(B^0_d \to \pi^+ \pi^-)\vert_{DE_L}$.
\par For the determination of $\alpha$ and the relative error, we define  
\be \sin 2\alpha^{eff}  = \frac{Im \lambda}{\vert \lambda \vert} 
\label{eq:saeff} \, ,\ee which is the quantity that can be extracted 
 from  the time-dependent asymmetry by measuring the coefficients of
$\cos (\Delta M_d t )$ and of $\sin (\Delta M_d t )$ in  eq.~(\ref{eq:asy}).
The uncertainty on the ``true" value of $\sin 2\alpha$ can be estimated  
by constructing 
\be \Delta = \sin 2 \alpha^{eff}  - \sin 2 \alpha  \, . \label{eq:delta} \ee
\par Finally we  present our
results for  the ratios $R_1=BR(B^0_d \to K^0 \bar K^0)/
BR(B^0_d \to \pi^+ \pi^-)\vert_{DE_L}$, $R_2=BR(B^+ \to \pi^+  K^0)/
BR(B^0_d \to \pi^+ \pi^-)\vert_{DE_L}$ and $R_3=BR(B^0_d \to K^+ \pi^-)/
BR(B^0_d \to \pi^+ \pi^-)\vert_{DE_L}$,  the values of which 
 have a strong dependence
on the contribution of charming-penguin diagrams.
\par In subsec.~\ref{subsec:one} the results  for 
$R$ and $R_{+-}$ under different assumptions on the 
operator matrix elements are presented; in subsec.~\ref{subsec:two} 
the effects  of  these assumptions  on the determination
of $\sin 2 \alpha$ are analyzed, while subsec.~\ref{subsec:three} contains
the numerical  results for $R_1$, $R_2$ and  $R_3$.
\subsection{Estimates of \boldmath$R$ and \boldmath$R_{+-}$}
\label{subsec:one}
\par  Given the large number of parameters, 
in order to understand which diagrams give important/unimportant
contributions, for  several values of $\xi$ and
$\delta_\xi$  we vary one single of the $\eta_i$s (and $\delta_L$)
 of the set defined  
in sec.~\ref{sec:me} at the time, while keeping  
all the others fixed to zero. We have also checked that the main
features of the results discussed below are the same irrespectively
of the choice of the  RP and of the renormalization scale $\mu$
($2$ GeV $\le \mu \le 10$ GeV).  
\begin{itemize} \item[a)] \underline{Colour suppression for $DE_L$ and
$CE_L$} Colour suppression depends on the value of the
Wilson coefficients, $\xi$ and the phase $\delta_\xi$. Information on the value
of $\xi$ is usually  obtained  by analyzing several $B$-decay channels 
($B \to D \pi$, $D \rho$, $D^* \pi$ and $D^* \rho$~\cite{xiref}) 
 and is correlated
to the Wilson  coefficients used in the analysis. In most of the 
cases, the leading order (LO) coefficients at $\mu=m_b$ are used and a value
of $\xi \sim 0.4$ is found. From a comparison of the LO
results, with those obtained at the next-to-leading order (NLO)
in different renormalization schemes (NDR, HV and RI~\cite{epp}), we find that 
the value of $\xi$ is not very sensitive to the (considered)
RP, whereas it can vary from
0.23, at  $\mu=10$ GeV, to 0.60, at $\mu=2$ GeV~\footnote{ This determination
of $\xi$ is only indicative, since it is not known whether the same
value of $\xi$ should be used for $B \to D \pi$ and $B \to \pi \pi$
decays.}.  The correlation between the  ratio  $C_1/C_2$ and the extracted
value of $\xi$ is shown in table \ref{tab:xixi}. We have taken
$a_2/a_1=0.25 \pm 0.05$, in agreement with most of the analyses of the
experimental data ~\cite{xiref}.
\begin{table}[t] \centering
\begin{tabular}{||c|c|c|c|c||}
\hline \hline
\multicolumn{1}{||c|}{}&
\multicolumn{4}{c||}{$C_1/C_2$}\\ \hline
$\mu$ (GeV)&
\multicolumn{1}{c|}{LO}&
\multicolumn{1}{c|}{NDR}&
\multicolumn{1}{c|}{HV}&
\multicolumn{1}{c||}{RI-Landau Gauge} \\
\hline 
2.0&-0.31&-0.27&-0.32&-0.24 \\
5.0&-0.19&-0.16&-0.19&-0.15\\
10.0&-0.12&-0.09&-0.12&-0.08\\
\hline\hline
\multicolumn{1}{||c|}{}&
\multicolumn{4}{c||}{$\xi$}\\ \hline
$\mu$ (GeV)&
\multicolumn{1}{c|}{LO}&
\multicolumn{1}{c|}{NDR}&
\multicolumn{1}{c|}{HV}&
\multicolumn{1}{c||}{RI-Landau Gauge} \\
\hline 
2.0&0.52(7)&0.49(8)&0.53(7)&0.46(8) \\
5.0&0.42(8)&0.39(9)&0.42(9)&0.39(9)\\
10.0&0.36(9)&0.33(9)&0.36(9)&0.32(9)\\
\hline\hline
\end{tabular}
\caption{\it{Values of $C_1/C_2$ and $\xi$ for different RP and at different values
of the renormalization scale $\mu$. The values of $\xi$ have been computed
assuming $a_2/a_1=0.25 \pm 0.05$, from $B \to D \pi$, $D \rho$, $D^* \pi$ and $D^* \rho$
decays, and using $\xi=(a_2/a_1-C_1/C_2)/(1-a_2/a_1 \times C_1/C_2)$.}}
\label{tab:xixi}
\end{table} 
In absence of a fully
consistent treatment of the amplitudes at the NLO, which could be
obtained if some  lattice  calculation   existed~\cite{epp}, we have no
 reason to prefer a particular value of the scale, or to maintain
 any correlation between $\mu$ and the value of $\xi$. 
 For this reason, in the following we take the coefficients
computed at the LO for $\mu=5$ GeV from ref.~\cite{epp}  but vary $\xi$ in the
range $0.2 \le \xi \le 0.5$, which covers almost all the values  in 
table~\ref{tab:xixi}. 
\par  We have a  comment on the choice of
the Wilson coefficients made in some recent studies which may be useful to 
the reader. In refs.~\cite{alek,desh} and \cite{paver} they used the
renormalization scheme independent coefficients introduced in 
ref.~\cite{buras2}, and computed for the full basis~(\ref{eq:basis})
in ref.~\cite{desh}. Though perfectly legitimate, this choice corresponds
to a value of the ratio $C_1/C_2=-0.27$ (and to $\xi \sim 0.47$), sensibly larger than those found
at leading order or at the NLO in the NDR, HV or RI schemes 
 at a renormalization scale $\mu \sim m_b$, see for example
ref.~\cite{epp}~\footnote{
Indeed the renormalization scheme of ref.~\cite{buras2}
has never been completely specified, because the external states on which 
the renormalization conditions were imposed  were not given explicitly in the
paper.}.
\par  On the basis of the discussion of the previous section,
 see eq.~(\ref{eq:xi00}), for $\delta_\xi=0$, the ratio $R$ has a minimum
for $\xi \sim -C_1/C_2 \sim 0.2$ and rises for larger values of $\xi$.
The suppression is softened  for non-vanishing values of $\delta_\xi$, as
can be seen in fig.~\ref{fig:xi},
and a value of  $R=0.03$--$0.06$ is obtained  for $\xi \sim 0.4$--$0.5$
and  $\delta_\xi=0.5$ (to be compared to $R=0.02$ at $\xi=0$). 
\begin{figure}[t]   
\begin{center}
\epsfxsize=0.7\textwidth
\epsfysize=0.5\textheight
\leavevmode\epsffile{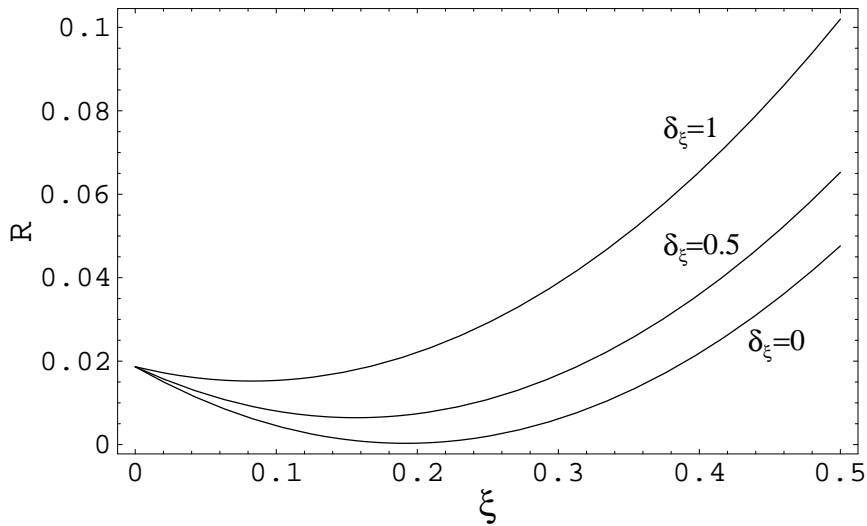}
\caption[]{\it{ The ratio $R$ is given as a function of $\xi$ at
three    values  of $\delta_\xi$. In the amplitudes, only  contributions
of  emission diagrams are included.}} 
\label{fig:xi}
\end{center}
\end{figure}
 Notice that $R_{+ -}$
only varies by less than $20$ \%  in the range
 of $\xi$ and $\delta_\xi$ considered here (it would
only  vary by about $10$ \%,  as a function of $\delta_\xi$, at
  $\xi=0.4$ fixed).
\item[b)] \underline{Annihilation and GIM-penguin diagrams} 
The effect of annihilation diagrams depends on $\delta_\xi$ and $\xi$. 
Let us first fix    $\xi=0.4$ and vary $\delta_\xi$. 
  For small values of $\delta_\xi$, up to $\delta_\xi \sim 0.4$,
$R$  increases with $\eta_A$ and reaches a value of 
 $\sim 0.1$ at $\eta_A =0.5$; for  larger values of $\delta_\xi$, 
 instead, there  is destructive  interference which 
balances the positive effect of $\delta_\xi$ on colour suppression.
The situation is illustrated in fig.~\ref{fig:etaa}.
\begin{figure}[t]   
\begin{center}
\epsfxsize=0.7\textwidth
\epsfysize=0.5\textheight
\leavevmode\epsffile{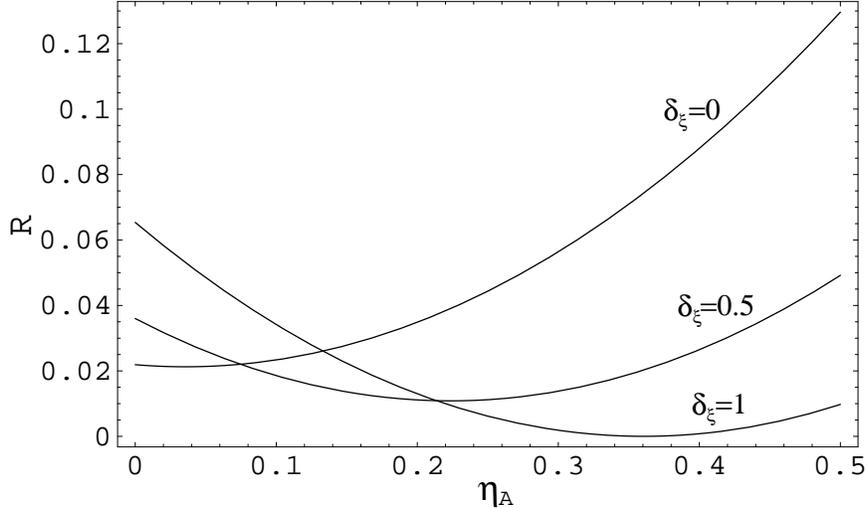}
\caption[]{\it{ The ratio $R$ is given as a function of $\eta_A$ at
three    values  of $\delta_\xi$. In the amplitudes, only contributions
of annihilation and emission diagrams are included.}} 
\label{fig:etaa}
\end{center}
\end{figure}
Contrary to  $R$ 
which is subject to appreciable variations, with  the same parameters
$R_{+ -}$ changes by less $20$ \% even at the largest value of $\eta_A$.
 Similar behaviours,
but on  different ranges of $\delta_\xi$ are observed  at other values of $\xi$.
\par The effect of GIM-penguin diagrams is essentially the same as for
annihilation diagrams, since, as can be immediately seen from 
eq.~(\ref{eq:daimp}),  the largest Wilson coefficient $C_2$
 multiplies the combination $\eta_A +\eta_P$.
\item[c)]\underline{$Q_5$ and $Q_6$ and Electro-penguins}  $Q_5$ and $Q_6$ 
give relatively small corrections both to $R$ and $R_{+ -}$,
because their coefficients are very small.
Of the two terms, the contribution of  $Q_5$, which has the smaller coefficient
and is colour suppressed, is always very small. $Q_6$ has the effect of 
changing by about $\sim 80$\%  and $\sim 5$\% the  ratios $R$ and $R_{+-}$
respectively when $\eta_6 = 2$. A large effect from $Q_6$
can only be obtained  at extremely large values of its matrix element.
In comparison, electro-penguin diagrams always give tiny corrections.
\item[d)] \underline{Charming penguins} The most important of all the
effects is given by charming  penguins. The explanation was already given in the
previous section: these diagrams correspond to the insertion of the
operators $Q_1$ and $Q_2$ which have large Wilson coefficients and 
 the contribution of which 
is also augmented by the factor $\tau$.  They may easily enhance 
the $B^0_d \to \pi^0 \pi^0$ amplitude, and   change also appreciably
the $B^0_d \to \pi^+ \pi^-$ rate. 
\begin{figure}[t]   
\begin{center}
\epsfxsize=0.7\textwidth
\epsfysize=0.5\textheight
\leavevmode\epsffile{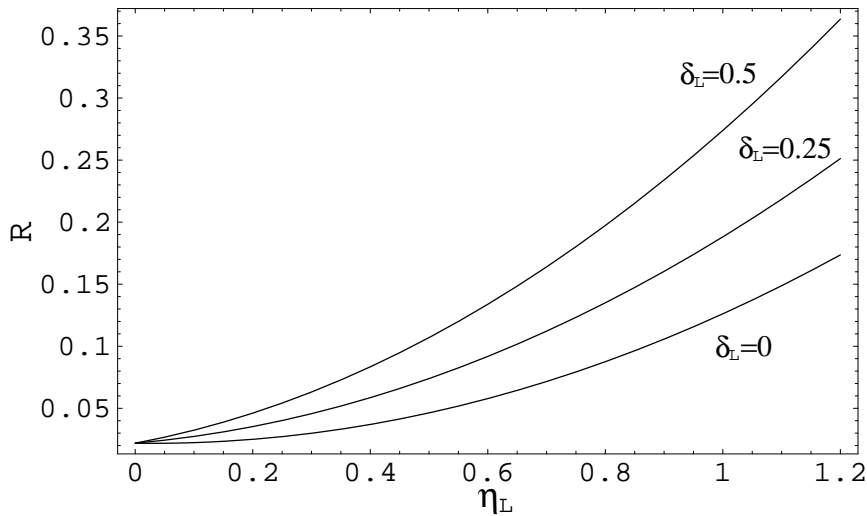}
\caption[]{\it{ The ratio $R$ is given as a function of $\eta_L$ at
for $\delta_L=0$, $0.25$ and $0.50$,
 at $\xi=0.4$ and $\delta_\xi=0$.
 In the amplitudes, only  contributions
of  emission diagrams and charming penguins are included.}} 
\label{fig:cpen}
\end{center}
\end{figure}
\begin{figure}[t]   
\begin{center}
\epsfxsize=0.7\textwidth
\epsfysize=0.5\textheight
\leavevmode\epsffile{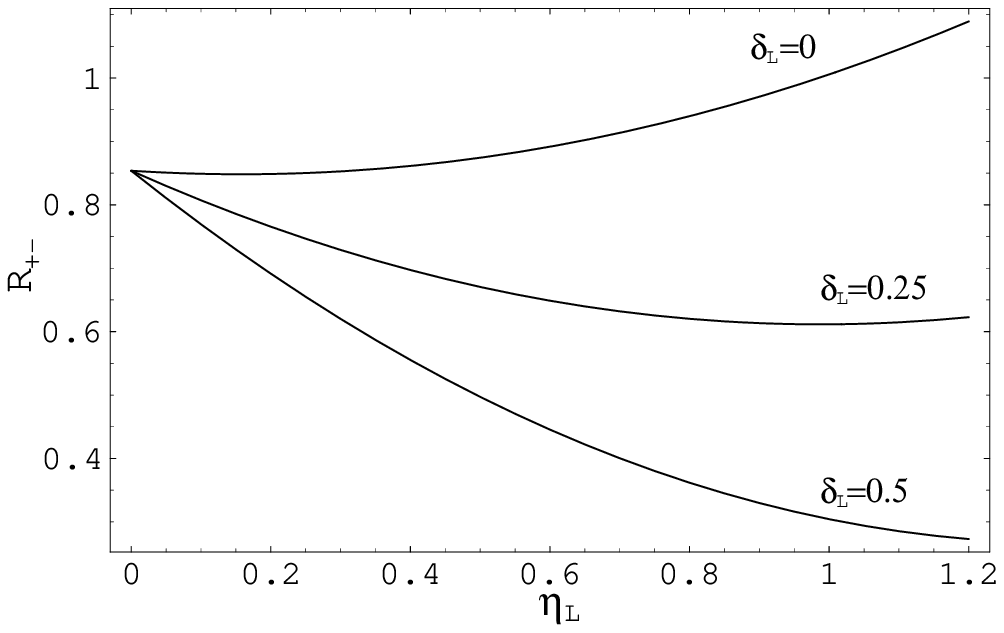}
\caption[]{\it{ The ratio $R_{+-}$ is given as a function of $\eta_L$ 
for $\delta_L=0$, $0.25$ and $0.50$, at $\xi=0.4$ and $\delta_\xi=0$.
 In the amplitudes, only  contributions
of  emission diagrams and charming  penguins are included.}} 
\label{fig:rpm}
\end{center}
\end{figure}

 In figs.~\ref{fig:cpen}~and~\ref{fig:rpm}, we give 
$R$ and $R_{+ -}$ as a function of $\eta_L$ at three
value of $\delta_L$ (for $\xi=0.4$ and $\delta_\xi=0$).
We observe that, for $\eta_L \sim 1.0$, there is a quite substantial effect
 when $\delta_L$ is different
from zero: $R$ can reach values as large as $0.25$ (for 
$\delta_L=0.25$) and even $0.35$ for 
$\delta_L=0.5$~\footnote{ The effect is striking for 
the $B^0_d \to \pi^+ \pi^-$ rate which at $\eta_L=1.2$
 and $\delta_L\sim 1.0$ almost vanishes.
 We believe, however, that this is a quite remote possibility.}.
 Even with $\delta_L=0$ we find  a large enhancement, which can lead
 to $R=0.10$--$0.15$ at relatively modest values of $\eta_L$. In this
 case there is even an increase of  about $20$\% of  $R_{+ -}$.
This discussion  shows that values of $BR(B^0_d \to \pi^0 \pi^0)$
as large as $1$--$3 \times 10^{-6}$ (assuming 
$BR(B^0_d \to \pi^+ \pi^-) = 1$--$2 \times 10^{-5}$) are indeed easily possible.
\end{itemize}
\subsection{Determination of \boldmath$\sin 2 \alpha$}
\label{subsec:two}
In this subsection, we discuss the uncertainties in the determination
of $\sin 2 \alpha$. These uncertainties are parametrized in terms
of the  shift  $\Delta$ introduced before.  $\Delta$ is
computed at different values of $\sin 2 \alpha$ which are obtained
by varying the CKM weak phase $\delta$. The values of
of $\sin 2 \alpha$  are computed from the expression  
$\sin 2 \alpha = Im(\tau/\tau^\star)$,
with $\tau$ given in the last of eqs.~(\ref{eq:wolf}), and using
$\rho=\sigma \cos \delta$ and $\eta=\sigma \sin \delta$ with  $\sigma=0.36$ 
and $-0.8 \le \cos \delta \le  +0.7$ ($\sin \delta \ge 0$).
In the figures below,  we give $\Delta$ as 
a function of  $\sin 2 \alpha^{eff}$
 because the latter is the quantity which is measured
experimentally. The strange behaviour of $\Delta$ for $\sin 2 \alpha^{eff}
\gtap 0.80$   comes from the fact that two different values of
$\cos \delta$, for $\cos \delta \ltap -0.6$, correspond to the same value of
$\sin 2 \alpha$. Had we varied $\cos \delta$ within the range
allowed by the combined analysis of the
 $K^0$--$\bar K^0$ and of the $B^0_d$--$\bar B^0_d$ mixing 
 amplitudes~\cite{ciuco},
i.e. $-0.3 \le \cos \delta \le 0.9$, the two-fold ambiguity 
would have disappeared, because the interval in $\cos \delta$ limits,
 in this case, $\sin 2 \alpha$ to values smaller than about $0.9$.
  \par In the following, 
we discuss  together the cases a)--c) of subsec.~\ref{subsec:one}
and the case d) separately.  
\begin{figure}[t]   
\begin{center}
\epsfxsize=0.7\textwidth
\epsfysize=0.5\textheight
\leavevmode\epsffile{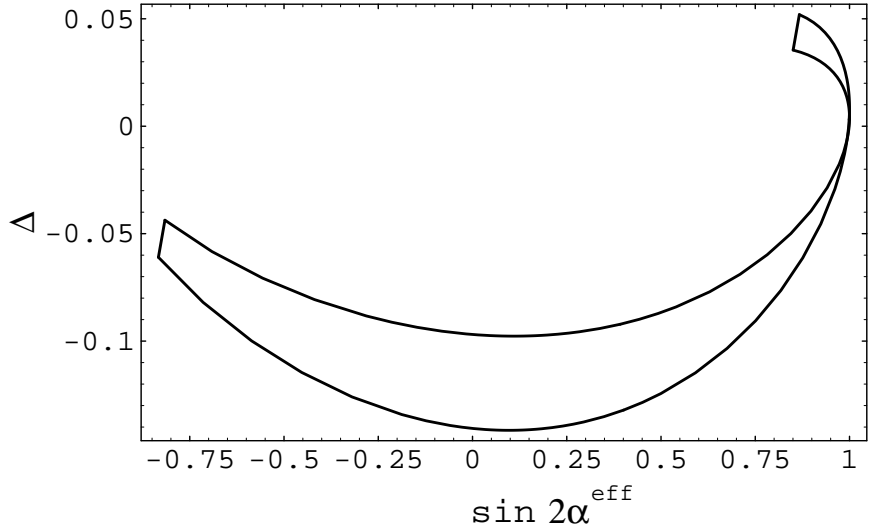}
\caption[]{\it{ A band of possible values for $\Delta$ as a function of 
$\sin 2 \alpha$ is given. This band corresponds to different choices of the parameters
introduced in 1)--5), see sec.~\protect\ref{sec:me}.}} 
\label{fig:family}
\end{center}
\end{figure}
\begin{itemize}\item[a)--c)]
In fig.~\ref{fig:family} we show a band of possible values for $\Delta$. 
This band
has been obtained in the following way:
a) by varying $\xi$ between $0.0$ and $0.5$
 with $\delta_\xi=0$, $0.5$ and $1.0$, and all the other $\eta_i$'s
taken to be zero; b) by fixing $\xi=0.4$ and 
varying $\eta_A$ between $0.0$ and $0.5$  with $\delta_\xi=0$, $0.5$ and $1.0$
($\eta_P=\eta_5=\eta_6=\eta_L=0$);
c) by taking $\xi=0.4$, $\delta_\xi=0$ and  $0.0 \le \eta_6 \le 2.0$
($\eta_A=\eta_P=\eta_5=\eta_L=0$). We see that for all the cases a)-c) 
the size  of $\Delta$   is about $-(0.05$--$0.15)$  in almost
the whole interval  of values of $\sin 2 \alpha$~\footnote{
 Smaller values are only found
when $\sin 2 \alpha$ is close to $+1$ or $-1$.}.  
This happens because the terms
depending on $\xi$, $\delta_\xi$, $\eta_A$ and $\eta_{5,6}$ give small variations
(of the order of $20$\% at most) to the $B^0_d \to \pi^+ \pi^-$ amplitude. 
These results are similar to those found in refs.~\cite{alek,paver} where
different hypotheses or approximations where used.
\begin{figure}[t]   
\begin{center}
\epsfxsize=0.7\textwidth
\epsfysize=0.5\textheight
\leavevmode\epsffile{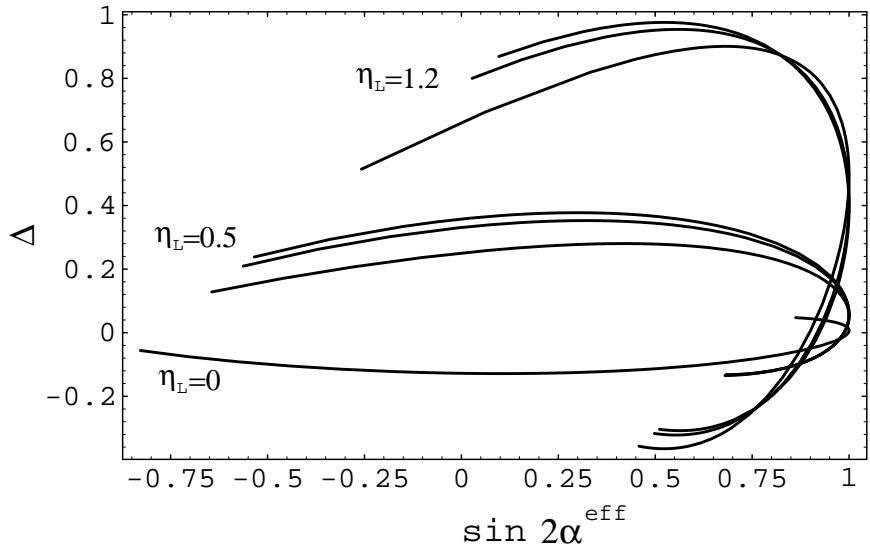}
\caption[]{\it{ A family of curves for $\Delta$ is given as a function of 
$\sin 2 \alpha$. They correspond to different choices of the parameters
$\eta_L$ ($\eta_L=0$, $0.5$, $1.2$) and $\delta_L$ 
($\delta_L=0$, $0.25$, $0.5$) introduced in d), see
sec.~\protect\ref{sec:me}.}} 
\label{fig:calpha}
\end{center}
\end{figure}
\begin{table}[t]\centering
\begin{tabular}{||c||c|c||c|c||}
\hline\hline
No. &
$\sin2\alpha^{eff}$ & $\Delta$ &
$\sin2\alpha^{eff}$ & $\Delta$\\
\hline\hline
& \multicolumn{2}{|c||}{$\cos\delta=-0.8$, $\sin 2\alpha=0.81$} &
\multicolumn{2}{|c||}{$\cos\delta=0$, $\sin 2\alpha=0.64$}\\
\hline
$2$ &
$0.86$ & $0.047$ &
$0.53$ & $-0.11$\\
$7$ &
$0.68$ & $-0.13$ &
$0.88$ & $0.24$\\
$8$ &
$0.68$ & $-0.13$ &
$0.88$ & $0.23$\\
$9$ &
$0.68$ & $-0.14$ &
$0.85$ & $0.21$\\
$14$ &
$0.73$ & $-0.087$ &
$0.8$ & $0.16$\\
$15$ &
$0.62$ & $-0.2$ &
$0.96$ & $0.31$\\
$18$ &
$0.72$ & $-0.09$ &
$0.82$ & $0.17$\\
$23$ &
$0.61$ & $-0.2$ &
$0.96$ & $0.32$\\
$27$ &
$0.65$ & $-0.17$ &
$0.92$ & $0.28$\\
\hline\hline
& \multicolumn{2}{|c||}{$\cos\delta=0.38$, $\sin 2\alpha=-0.04$} &
\multicolumn{2}{|c||}{$\cos\delta=0.7$, $\sin 2\alpha=-0.77$}\\
\hline
$2$ &
$-0.16$ & $-0.12$ &
$-0.83$ & $-0.056$\\
$7$ &
$0.34$ & $0.38$ &
$-0.53$ & $0.24$\\
$8$ &
$0.32$ & $0.35$ &
$-0.56$ & $0.21$\\
$9$ &
$0.24$ & $0.27$ &
$-0.64$ & $0.13$\\
$14$ &
$0.17$ & $0.21$ &
$-0.67$ & $0.11$\\
$15$ &
$0.53$ & $0.57$ &
$-0.36$ & $0.42$\\
$18$ &
$0.21$ & $0.25$ &
$-0.63$ & $0.14$\\
$23$ &
$0.52$ & $0.56$ &
$-0.38$ & $0.39$\\
$27$ &
$0.43$ & $0.47$ &
$-0.47$ & $0.3$\\
\hline\hline
\end{tabular}
\caption{\it{We give results for $\sin 2 \alpha^{eff}$ and $\Delta$
obtained with different  choices of the parameters  $\delta_\xi$, 
$\dots$, $\eta_i$s. In all cases $\xi=0.4$. In the table, the
value of $\cos \delta$ used in the different examples, 
and the corresponding value
of   $\sin \alpha$, are also shown. The label ``No.", 
which is the same of table~\protect\ref{tab:Rs},  allows
a study of the correlation between $\Delta$ and the values of the
branching ratios discussed in subsec.~\protect\ref{subsec:three}.}}
\label{tab:alpha}
\end{table} 
\item[d)] We varied $\eta_L$ and $\delta_L$ as at the point d) of subsection
\ref{subsec:one}, while taking $\xi=0.4$ and $\delta_\xi$ and
all the other $\eta_i$s to be zero. In this case
the uncertainty  on $\sin 2 \alpha$ can be very large, corresponding to
$\Delta \sim 0.4$--$0.8$, see fig.~\ref{fig:calpha}. The reason, as already discussed above,
is that charming  penguins give  large corrections 
to the  $B^0_d \to \pi^+ \pi^-$ amplitude too, while in the other cases considered
above [a)--c)], the
corrections  are large only  for the  $B^0_d \to \pi^0 \pi^0$ amplitude
where large cancellations occur.
\end{itemize}
\par In order to guide the reader, we end this subsection by giving 
in table \ref{tab:alpha} a set of numerical  results for
$\sin 2 \alpha^{eff}$ and $\Delta$ obtained  for different  choices
of the parameters introduced in a)--d) (also reported in the table).
In the set of examples given in table~\ref{tab:alpha}, $\Delta \ltap
0.6$.
\subsection{Charming-penguin dominated branching ratios}
\label{subsec:three}
Our main results for the ratios of rates introduced before are summarized
in table \ref{tab:Rs}, where a large set of possibilities
has been considered. We have introduced a label ``No." to distinguish
the different cases which will be discussed below~\footnote{
In order to check the stability of the results,
 other cases have been analyzed but not given in table 
\ref{tab:Rs}.}.
The table is instructive because it shows the strong correlation
between  $R$  and   $R_{1,2,3}$,  particularly  when 
 charming-penguin contributions are large.  We will make use
of this correlation, combined to the experimental information
$BR(B^0_d \to K^+ \pi^-)= (1.5 ^{+0.5}_{-0.4} \pm 0.2 ) \times 10^{-5}$
and $BR(B^0_d \to \pi^+ \pi^-)  < 1.5 \times 10^{-5}$~\cite{CLEO},
and to prejudices based on theoretical estimates of the matrix elements,
in order to reduce the uncertainties. 
Given the present precision 
in the measurement of  $BR(B^0_d \to K^+ \pi^-)$,
and our  ignorance on the matrix elements, the discussion below
is, at the moment,
preliminary and semi-quantitative  and only  intended to illustrate a method
which may become very useful when the experimental precision will increase
and   theoretical predictions for  GIM- and charming-penguin diagrams
will become available. 
\begin{table}[t]\centering
\begin{tabular}{||c|c|c|c|c|c|c||c|c|c|c|c|c||}
\hline\hline
No. & $\xi$ & $\delta_\xi$ & $\eta_{A,P}$ & $\eta_{5,6}$ & $\eta_L$ &
$\delta_L$ & ${\cal R}$ & ${\cal R}_{+-}$ & ${\cal R}_1$ &
${\cal R}_2$ & ${\cal R}_3$\\
\hline
$1$ & $0$ & $0$ & $0$ & $0$ & $0$ & $0$ &
$0.019$ & $1.$ & $0.0044$ & $0.11$ & $0.083$\\
$2$ & $0.4$ & $0$ & $0$ & $0$ & $0$ & $0$ &
$0.022$ & $0.85$ & $0.0025$ & $0.061$ & $0.076$\\
$3$ & $0.4$ & $0.5$ & $0$ & $0$ & $0$ & $0$ &
$0.036$ & $0.87$ & $0.0028$ & $0.067$ & $0.075$\\
$4$ & $0.4$ & $0$ & $0.25$ & $0$ & $0$ & $0$ &
$0.13$ & $1.1$ & $0.04$ & $0.083$ & $0.13$\\
$5$ & $0.4$ & $0$ & $0$ & $1$ & $0$ & $0$ &
$0.025$ & $0.87$ & $0.013$ & $0.31$ & $0.3$\\
$6$ & $0.4$ & $0$ & $0$ & $2$ & $0$ & $0$ &
$0.032$ & $0.89$ & $0.031$ & $0.75$ & $0.71$\\
$7$ & $0.4$ & $0$ & $0$ & $0$ & $0.5$ & $0$ &
$0.046$ & $0.87$ & $0.034$ & $0.82$ & $0.94$\\
$8$ & $0.4$ & $0$ & $0$ & $0$ & $0.5$ & $0.25$ &
$0.074$ & $0.67$ & $0.042$ & $1.$ & $1.3$\\
$9$ & $0.4$ & $0$ & $0$ & $0$ & $0.5$ & $0.5$ &
$0.11$ & $0.5$ & $0.066$ & $1.6$ & $2.1$\\
$10$ & $0.4$ & $0$ & $0$ & $0$ & $1$ & $0$ &
$0.13$ & $1.$ & $0.18$ & $4.3$ & $4.5$\\
$11$ & $0.4$ & $0$ & $0$ & $0$ & $1$ & $0.25$ &
$0.19$ & $0.61$ & $0.21$ & $5.$ & $5.6$\\
$12$ & $0.4$ & $0$ & $0$ & $0$ & $1$ & $0.5$ &
$0.27$ & $0.3$ & $0.29$ & $7.1$ & $8.1$\\
$13$ & $0.4$ & $0.5$ & $0$ & $1$ & $0.5$ & $0$ &
$0.016$ & $0.86$ & $0.014$ & $0.34$ & $0.46$\\
$14$ & $0.4$ & $0.5$ & $0$ & $1$ & $0.5$ & $0.25$ &
$0.043$ & $0.66$ & $0.027$ & $0.65$ & $0.93$\\
$15$ & $0.4$ & $0.5$ & $0$ & $2$ & $1$ & $0$ &
$0.025$ & $0.9$ & $0.084$ & $2.$ & $2.3$\\
$16$ & $0.4$ & $0.5$ & $0$ & $2$ & $1$ & $0.25$ &
$0.089$ & $0.52$ & $0.13$ & $3.1$ & $3.7$\\
$17$ & $0.4$ & $0.5$ & $0.25$ & $0$ & $0.5$ & $0$ &
$0.15$ & $1.3$ & $0.19$ & $0.74$ & $0.89$\\
$18$ & $0.4$ & $0.5$ & $0.5$ & $0$ & $0.5$ & $0.25$ &
$0.54$ & $2.$ & $0.45$ & $0.98$ & $1.4$\\
$19$ & $0.4$ & $0.5$ & $0.25$ & $0$ & $1$ & $0$ &
$0.3$ & $1.7$ & $0.45$ & $4.$ & $4.3$\\
$20$ & $0.4$ & $0.5$ & $0.5$ & $0$ & $1$ & $0.25$ &
$0.82$ & $2.3$ & $0.82$ & $4.8$ & $5.6$\\
$21$ & $0.4$ & $0.5$ & $0.25$ & $1$ & $0.5$ & $0$ &
$0.12$ & $1.2$ & $0.14$ & $0.3$ & $0.43$\\
$22$ & $0.4$ & $0.5$ & $0.5$ & $1$ & $0.5$ & $0.25$ &
$0.47$ & $1.8$ & $0.37$ & $0.63$ & $1.1$\\
$23$ & $0.4$ & $0.5$ & $0.25$ & $1$ & $1$ & $0$ &
$0.25$ & $1.5$ & $0.36$ & $2.9$ & $3.1$\\
$24$ & $0.4$ & $0.5$ & $0.5$ & $1$ & $1$ & $0.25$ &
$0.74$ & $2.2$ & $0.71$ & $3.8$ & $4.6$\\
$25$ & $0.4$ & $0.5$ & $0.25$ & $2$ & $0.5$ & $0$ &
$0.09$ & $1.1$ & $0.093$ & $0.061$ & $0.17$\\
$26$ & $0.4$ & $0.5$ & $0.5$ & $2$ & $0.5$ & $0.25$ &
$0.42$ & $1.7$ & $0.3$ & $0.48$ & $0.9$\\
$27$ & $0.4$ & $0.5$ & $0.25$ & $2$ & $1$ & $0$ &
$0.21$ & $1.4$ & $0.29$ & $1.9$ & $2.1$\\
$28$ & $0.4$ & $0.5$ & $0.5$ & $2$ & $1$ & $0.25$ &
$0.67$ & $2.$ & $0.61$ & $3.$ & $3.8$\\
\hline\hline
\end{tabular}
\caption{\it{We give results for $R=BR(B^0_d \to \pi^0 \pi^0)/
BR(B^0_d \to \pi^+ \pi^-)\vert_{DE_L}$,
$R_{+ -}=BR(B^0_d \to \pi^+ \pi^-)/
BR(B^0_d \to \pi^+ \pi^-)\vert_{DE_L}$,  $R_1=BR(B^0_d \to K^0 \bar K^0)/
BR(B^0_d \to \pi^+ \pi^-)\vert_{DE_L}$, $R_2=BR(B^+ \to \pi^+  K^0)/
BR(B^0_d \to \pi^+ \pi^-)\vert_{DE_L}$ and $R_3=BR(B^0_d \to K^+ \pi^-)/
BR(B^0_d \to \pi^+ \pi^-)\vert_{DE_L}$, 
obtained with different choices of the parameters $\xi$, $\dots$, $\eta_i$s.}}
\label{tab:Rs}
\end{table} 
\par We proceed as follows:
\begin{itemize}
\item[i)] \underline{$BR(B^0_d \to \pi^+
\pi^-)$} Different theoretical estimates of this branching ratio, 
obtained by
using the factorization hypothesis  but with different models to evaluate
the matrix elements, are consistent within a factor of two and give   
$BR(B^0_d \to \pi^+\pi^-)=1$--$2 \times 10^{-5}$~\cite{dibart}--\cite{alek}.
The spread of these predictions is mostly due to  differences in  the  values 
of the   form factors of the weak-currents and in the value of  $\xi$
used for the calculation of the emission diagrams.
The results, instead,   are only    marginally affected by the
contribution of the  penguin operators
$Q_3$--$Q_{10}$, which in some cases have been omitted
(charming-penguin diagrams have never been considered). This happens
for the following reason:
in the factorized case the matrix elements
 of $Q_5$ and $Q_6$, which are expected
to give the largest contributions of all  penguin operators,
correspond  to $\eta_{5,6}\sim 1$ for which
the correction is
of a few percent, as can been seen by comparing No. 2 to No. 5
in table \ref{tab:Rs}.  This explains the relative stability of
the theoretical results. We then 
assume that $1$--$2 \times 10^{-5}$ is a range
representative of the theoretical uncertainty for
emission diagrams~\footnote{ 
Predictions  on $BR(B^0_d \to \pi^0\pi^0)$ are, instead, very unstable
because in this case  even a small
error on the evaluation of a single diagram is amplified by
the large cancellations of the different contributions.
This is also shown by comparing different entries in  the table, 
where variations of $R$  over  one order of magnitude are  observed.}.
On the basis of the above discussion, in the following we  consider
three possibilities $BR(B^0_d \to \pi^+ \pi^-)|_{DE+CE}=(0.5,\,
1.0, \, 1.5)\times 10^{-5}$. The latter two cases 
have been used because they are  in the ballpark of all estimates, 
the first value
because it accounts for  large deviations from the factorization 
predictions.  Values of  
$BR(B^0_d \to \pi^+ \pi^-)|_{DE+CE}$ larger than 
$2 \times 10^{-5}$ are excluded on the basis of ii) below.
$BR(B^0_d \to \pi^+ \pi^-)|_{DE+CE}$ corresponds to the case No. 2 of
table \ref{tab:Rs}, where $R_{+-}=0.85$. Thus, for
example,   if we assume
$BR(B^0_d \to \pi^+ \pi^-)|_{DE+CE}= 1.5 \times 10^{-5}$, the ratio
$R=0.54$ (No. 18) gives $BR(B^0_d \to \pi^0 \pi^0)= 0.54/0.85 \times 1.5
\times 10^{-5}= 0.95 \times 10^{-5}$. 
\item[ii)]\underline{$BR(B^0_d \to K^+ \pi^-)$ vs
$BR(B^0_d \to \pi^+ \pi^-)$}
From CLEO we expect  $BR(B^0_d \to K^+ \pi^-)$  to be  larger than
$BR(B^0_d \to \pi^+ \pi^-)$. Thus we  exclude all those cases
where $BR(B^0_d \to K^+ \pi^-) \le BR(B^0_d \to \pi^+ \pi^-)$,
namely No. 1--6, 13, 17, 18, 21, 22, 25, 26.
\item[iii)]\underline{$BR(B^0_d \to K^+ \pi^-)$}
We also exclude all cases where  
$BR(B^0_d \to K^+ \pi^-) \le 1 \times 10^{-5}$
or  $BR(B^0_d \to K^+ \pi^-) \ge 2 \times 10^{-5}$.
\end{itemize}
Indeed, from the experimental results, $BR(B^0_d \to K^+ \pi^-) \sim
2.5\times 10^{-5}$ or $BR(B^0_d \to K^+ \pi^-) \sim
0.5 \times 10^{-5}$ and also $BR(B^0_d \to K^+ \pi^-)
 \sim  BR(B^0_d \to \pi^+ \pi^-)$ are still open possibilities. However,
 we want   to stretch here the experimental constraints
in order  to show how this analysis works.
\par We now discuss  the results of table \ref{tab:BRs},
where the main results  of the selection based on i)--iii)
are given.
For comparison, in this table we also give the results for No. 2,
where only emission diagrams contribute.
\begin{table}[t]\centering
\begin{tabular}{||c||c|c|c|c|c||}
\hline \hline
No. & $B^0_d \to \pi^+ \pi^-$ &
 $B^0_d \to \pi^0 \pi^0$ & $B^0_d \to K^0 \bar K^0$ &
$B^+ \to \pi^+ K^0$ & $B^0_d \to K^+ \pi^-$\\
& $BR\times 10^{5}$ & $BR\times 10^{5}$ & $BR\times 10^{5}$ &
$BR\times 10^{5}$& $BR\times 10^{5}$ \\
\hline\hline
\multicolumn{6}{||c||}
{$BR(B^0_d \to \pi^+ \pi^-)_{DE+CE} =1.5\times 10^{-5}$}\\
\hline
$2$ & $1.5$& $0.05$ & $0.004$ & $0.11$ & $0.13$ \\ \hline
$7$ & $1.5$ &  $0.08$  & $0.06$ & $1.4$ & $1.7$\\
$14$ & $1.2$ & $0.08$ &  $0.05$ & $1.1$ & $1.6$\\
\hline\hline
\multicolumn{6}{||c||}
{$BR(B^0_d \to \pi^+ \pi^-)_{DE+CE} =1\times 10^{-5}$}\\
\hline
$2$ & $1$ & $0.03$ & $0.003$ & $0.07$ & $0.09$ \\ \hline
$7$ & $1$ &  $0.05$  & $0.04$ & $1$ & $1.1$\\
$8$ & $0.8$ & $0.09$ &  $0.05$ & $1.2$ & $1.5$\\
$14$ & $0.8$ & $0.05$ &  $0.03$ & $0.8$ & $1.1$\\
\hline\hline
\multicolumn{6}{||c||}
{$BR(B^0_d \to \pi^+ \pi^-)_{DE+CE}=0.5\times 10^{-5}$}\\
\hline 
$2$ & $0.5$& $0.01$ & $0.001$ & $0.04$ & $0.04$ \\ \hline
$9$ & $0.3$ & $0.06$ &  $0.04$ & $0.9$ & $1.2$\\
$15$ & $0.5$ & $0.01$ &  $0.05$ & $1.2$ & $1.4$\\
$23$ & $0.9$ & $0.1$ &  $0.2$ & $1.7$ & $1.8$\\
$27$ & $0.8$ & $0.1$ &  $0.2$ & $1.1$ & $1.2$\\
\hline\hline
\end{tabular}
\caption{\it{Branching ratios for $B^0_d \to \pi^+ \pi^-$,
 $B^0_d \to \pi^0 \pi^0$, $B^0_d \to K^0 \bar K^0$,
$B^+ \to \pi^+ K^0$ and $B^0_d \to K^+ \pi^-$. The cases are labeled as
in table \protect\ref{tab:Rs}. The $BR$s are normalized assuming three
different values of $BR(B^0_d \to \pi^+ \pi^-)|_{DE+CE}$.}}
\label{tab:BRs}
\end{table} 
We first notice that the 
charming-penguin enhancement of $BR(B^+ \to \pi^+ K^0)$ and 
$BR(B^0_d \to K^+ \pi^-)$ is always very large, corresponding
to a factor of 10--40, with respect to standard case No. 2.
A stable prediction of the results in
the table is that we also expect  $BR(B^+ \to \pi^+ K^0)$ to be comparable to 
$BR(B^0_d \to K^+ \pi^-)$ and of the order $1 \times 10^{-5}$.
Although,  on the basis of table
\ref{tab:Rs},  a much large enhancement of $BR(B^0_d \to \pi^0  \pi^0)$
would be possible,
 the constraints i)--iii) limit the effect of charming  penguins
so that the final result is about a factor of 2  larger than previous 
estimates~\cite{dibart,altri}. A larger enhancement is still possible,
however,
if we relax the selection constraints. For example, without ii), 
case No. 18 is acceptable and, for $BR(B^0_d \to
\pi^+ \pi^-)|_{DE+CE}=0.5 \times 10^{-5}$, it gives
\bea BR(B^0_d \to \pi^+ \pi^-)&=& 1.2  \times 10^{-5}\, ,
\,\,\,\,\,\,\,\,\,\, BR(B^0_d \to \pi^0 \pi^0)=0.3 \times 10^{-5}\, , \nn
\\ BR(B^0_d \to K^0 \bar K^0)&=& 0.3 \times 10^{-5} \, ,
\,\,\,\,\,\,\,\,\,\, BR(B^+ \to \pi^+ K^0)=0.6 \times 10^{-5}\, , \nn
\\ BR(B^0_d \to K^+ \pi^-)&=&0.8 \times 10^{-5}\, .
  \label{eq:exa} \eea
These  results give a large $BR(B^0_d \to \pi^0 \pi^0)$, and are
still compatible with the experimental measurement of   
$BR(B^0_d \to K^+ \pi^-)$ and the present limit on $BR(B^0_d \to \pi^+ \pi^-)$.
This example also show the importance of the reduction of the experimental
error for determining charming-penguin effects.
\par We finally notice that 
for $BR(B^0_d \to K^0 \bar K^0)$, besides charming penguins,
GIM-penguins are also   relatively  important. Moreover
GIM- and  charming-penguin diagrams appear in  
the r.h.s. of  eq.~(\ref{eq:apkkb}) in   the same combination
as  in $A^t_0$, eq.~(\ref{eq:mostros}). A  measurement of 
$BR(B^0_d \to K^0 \bar K^0)$ is then  
very important  for the determination of the
combined  value of  GIM and charming-penguin diagrams in $B^0_d \to \pi \pi$
decays and may lead to a reduction of the uncertainties in the
determination of $\sin 2 \alpha$.
Unfortunately, in most of the cases,
this branching ratio remains rather small, 
$BR(B^0_d \to K^0 \bar K^0)\sim 5 \times 10^{-7}$,
in spite of the enhancement due to charming  penguins.
For this reason this  may be a difficult measurement  even for
$B$-factories.
However, from the example discussed in (\ref{eq:exa}) and the results
of the table,
a $BR(B^0_d \to K^0 \bar K^0)=2$--$3\times 10^{-6}$ is still
compatible with the present experimental constraints and the experimental
search of this decay mode is very important.
 
\section{Conclusions}
In this paper  we have discussed  several effects which could  enhance the
$B^0_d \to  \pi^0 \pi^0$ branching ratio.
Among these, we have shown  the presence of diagrams
involving operators containing charmed  quarks, denoted as 
``charming  penguins", which were never studied before.
Since there is no reason {\it a priori} to expect the value of these
diagrams  to be small, and the corresponding Wilson coefficients are
of $O(1)$, and not of $O(\alpha_s /12 \pi \ln (m_t^2/\mu^2))$,
they may have a large effect in the $B^0_d \to \pi^0 \pi^0$ decay rate and
in  the determination of the weak phase $\alpha$ extracted from the measurement
of the  $B^0_d \to \pi^+ \pi^-$ time-dependent asymmetry.  
 In absence of any realistic estimate of the value of 
charming-penguin diagrams, we allow them to vary within  reasonable
 ranges of values and show that, without further constraints,
they can lead to a 
 $BR(B^0_d \to \pi^0 \pi^0 )\sim 1$--$3 \times 10^{-6}$. Correspondingly,
the uncertainty in determination of $\sin 2\alpha$ from the
time-dependent asymmetry could be as large as $0.4$--$0.8$. \par  
We have also shown that GIM- and charming-penguin diagrams
dominate the $B^0_d \to K^0 \bar K^0$  amplitude  because 
emission and annihilation  diagrams are only produced  by the insertion  of
the operators  $Q_3$--$Q_{10}$ which have rather small Wilson coefficients.
GIM- and charming-penguin diagrams appear in the 
$B^0_d \to K^0 \bar K^0$  amplitude in the same combination as in the
$B^0_d \to \pi \pi$ case. Thus, if we assume
$SU(3)$ symmetry,  a measurement of $BR(B^0_d \to
K^0 \bar K^0)$ contains important information on these contributions and
may help to reduce the uncertainty in the extraction of $\sin 2\alpha$ from
the $B^0_d \to \pi^+ \pi^-$ time-dependent asymmetry.
\par  A striking effect of charming-penguins occurs in
$B^+ \to \pi^+ K^0$ and  $B^0_d \to K^+ \pi^-$ decays.
In this case, the enhancement
easily gives values of
 $BR(B^+ \to \pi^+ K^0)$ and $BR(B^0_d \to K^+ \pi^-)$ of 
about $1 \times 10^{-5}$,
 $10$--$40$  times  larger than those obtained by considering 
emission diagrams only. This is because in the latter case the main 
contribution is Cabibbo suppressed. 
Our findings are supported by  the recent
results of the CLEO Collaboration~\cite{CLEO}. 
Theoretical predictions of the charming-penguin (and GIM-penguin) 
amplitudes, either
from the lattice, or with QCD sum rules, or with any other non-perturbative
 technique,  are then  urgently needed. 
\section*{Acknowlegments}
We thank F.~Ferroni,
M.~Giorgi, M.~Gronau, A.~Masiero, O.~P\`ene, R.~Poling  and
A.~Pugliese for very useful discussions.
We acknowledge the partial support by the EC contract CHRX-CT93-0132.
G.M. acknowledges the  Phys. Dept. of the Universidad Aut\'onoma de Madrid
for its hospitality and the Fundacion IBERDROLA for its support. 
We acknowledge partial support by M.U.R.S.T. 

\end{document}